\documentclass[lettersize,journal]{IEEEtran}
\usepackage{amsmath,amsfonts}
\usepackage{algorithmic}
\usepackage{algorithm}
\usepackage{array}
\usepackage[caption=false,font=normalsize,labelfont=small,textfont=sf]{subfig}
\usepackage{textcomp}
\usepackage{stfloats}
\usepackage{url}
\usepackage{verbatim}
\usepackage{graphicx}
\usepackage{enumitem}
\usepackage{xurl}
\usepackage{pifont}
\usepackage{svg}
\graphicspath{ {./images/} }
\usepackage{cite}
\usepackage{fancyhdr}
\usepackage{lipsum}
\begin{document}
\title{Edge Video Analytics: A Survey on Applications, Systems and Enabling Techniques}

\author{Renjie Xu,~\IEEEmembership{Student Member,~IEEE,} Saiedeh Razavi,~\IEEEmembership{Member,~IEEE,} Rong Zheng,~\IEEEmembership{Senior Member,~IEEE}}

\markboth{}%
{Shell \MakeLowercase{\textit{et al.}}: A Sample Article Using IEEEtran.cls for IEEE Journals}


\maketitle

\thispagestyle{fancy}

\begin{abstract}
Video, as a key driver in the global explosion of digital information, can create tremendous benefits for human society. Governments and enterprises are deploying innumerable cameras for a variety of applications, e.g., law enforcement, emergency management, traffic control, and security surveillance, all facilitated by video analytics (VA). This trend is spurred by the rapid advancement of deep learning (DL), which enables more precise models for object classification, detection, and tracking. Meanwhile, with the proliferation of Internet-connected devices, massive amounts of data are generated daily, overwhelming the cloud. Edge computing, an emerging paradigm that moves workloads and services from the network core to the network edge, has been widely recognized as a promising solution. The resulting new intersection, edge video analytics (EVA), begins to attract widespread attention. Nevertheless, only a few loosely-related surveys exist on this topic. The basic concepts of EVA (e.g., definition, architectures) were not fully elucidated due to the rapid development of this domain. To fill these gaps, we provide a comprehensive survey of the recent efforts on EVA. In this paper, we first review the fundamentals of edge computing, followed by an overview of VA. EVA systems and their enabling techniques are discussed next. In addition, we introduce prevalent frameworks and datasets to aid future researchers in the development of EVA systems. Finally, we discuss existing challenges and foresee future research directions. We believe this survey will help readers comprehend the relationship between VA and edge computing, and spark new ideas on EVA.
\end{abstract}


\begin{figure*}[htbp]
\centering
\includegraphics[width=0.95\textwidth]{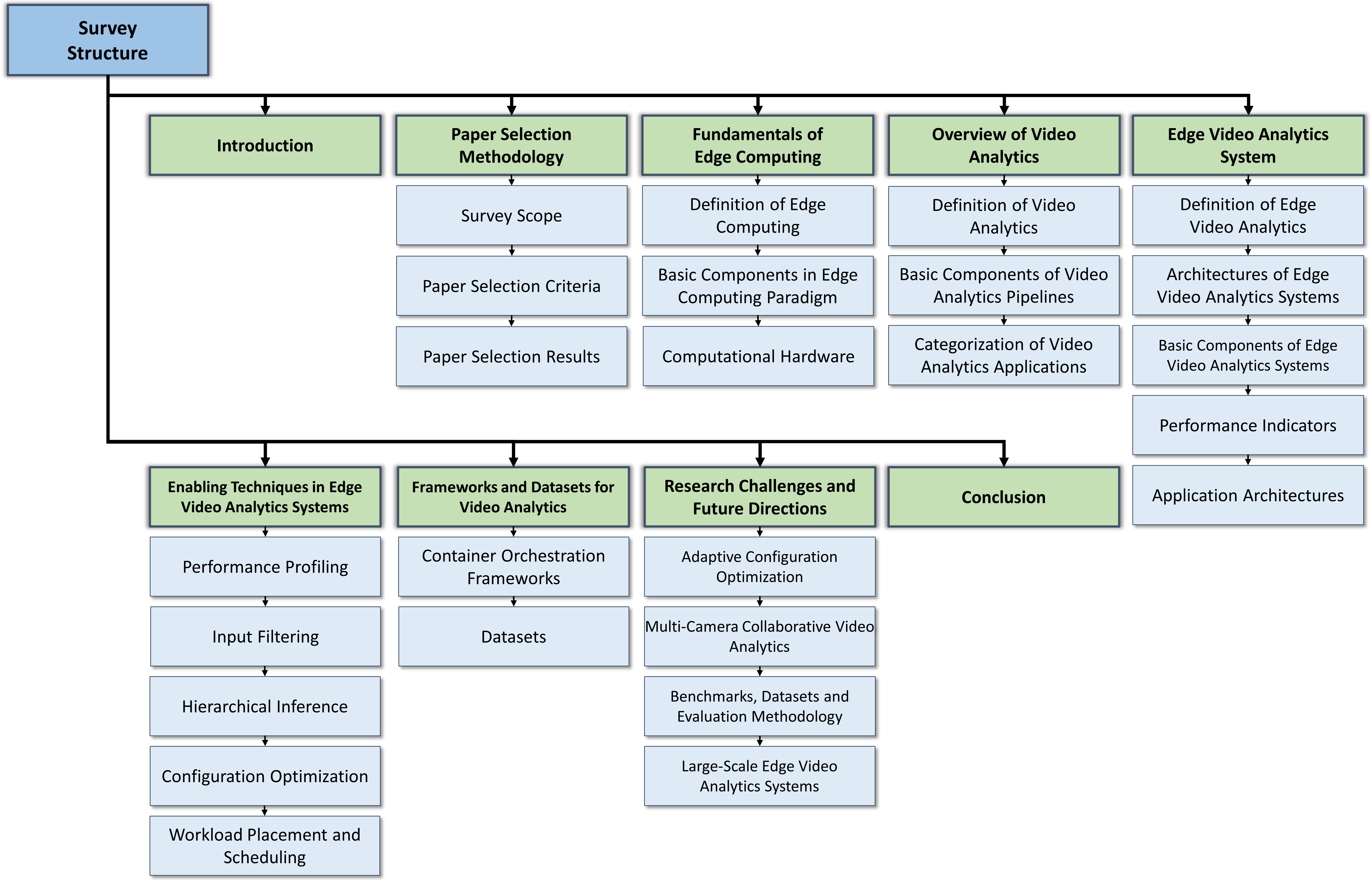}
\caption{Organization of this survey.}
\label{structure}
\end{figure*}

\section{Introduction}
\IEEEPARstart{C}{ameras} are in every corner of our cities in this information-centric era. According to a 2019 report by the Information Handling Services (IHS), one surveillance camera is installed for every 8 people on the planet nowadays, with mature markets (e.g., China and the United States) having one camera for every 4 people \cite{IHS}. As predicted by LDV Capital, the number of cameras (including various types of cameras) in the world will reach 45 billion by 2022, experiencing an increase of 216\% in the past five years \cite{LDV}. This trend poses a great challenge for humans to discover useful information from explosive video data. It is beyond human capacity to make sense of what is happening in all video feeds manually. Video analytics (VA) is a technique that can automatically and efficiently recognize objects and identify interesting events in unstructured video data. It can drive a large number of applications with wide-ranging impacts on our society. Examples of such applications include security surveillance in public and private venues, assisted and autonomous driving and consumer applications, including digital assistants for real-time decision-making.

Early-stage video analytics is based on conventional image processing techniques, which mainly rely on human expertise and empirical knowledge, and thus are not robust to changes in lighting conditions, viewing angles, weather conditions, etc. Deep learning (DL), as a research hot spot in the past decades, has made striking breakthroughs in many fields, especially in computer vision (CV). Advanced CV technologies, e.g., object classification, detection, and tracking, enable extracting more accurate information and insights from video feeds. The resulting insights can help people make smarter and faster decisions.

However, many DL-driven applications are compute-intensive, thus not friendly to resource-constrained Internet-of-Things (IoT) devices. The conventional wisdom is to offload all workloads from devices to a cloud via wide area networks (WANs), where powerful data centers are located. This computing paradigm, known as cloud computing, suffers from high service delays due to long geographical distances and potential network congestion. According to a report by International Data Corporation (IDC), worldwide data will reach 175 zettabytes (ZB) by 2025, 51\% of which will be created by IoT devices \cite{IDC}. Digesting such massive data in the cloud incurs excessive delays, making such solution inadequate for mission-critical applications, e.g., security surveillance \cite{SurveilEdge} and autonomous driving \cite{VIPS}, where the safety of citizens and customers can be compromised if responses arrive too late.

Edge computing, a rising computing paradigm, has recently been recognized as a viable alternative to cloud computing. It is a distributed architecture that reduces latency by hosting applications and computing resources at locations geographically closer to the data source. Simply put, edge computing alleviates data transfer latency by processing data on local edge nodes rather than in a remote cloud. An edge node can vary in size, ranging from tiny processing units co-located with IoT devices, to IT infrastructures in the physical proximity of base stations (BSs). These nodes, distributed at the network edge, can significantly alleviate the workloads and traffic congestions of the cloud, thereby reducing the service delay and improving the quality of experience (QoE) of users.

Obviously, edge computing is an extension of cloud computing by pushing centralized workloads to the network edge. Instead of entirely relying on the cloud, edge computing, a flexible computing paradigm leveraging both edge and cloud capabilities effectively, is gaining traction in building VA systems. Therefore, we are now witnessing the convergence of video analytics and edge computing, namely, \textit{edge video analytics} (EVA). Major service providers, e.g., Microsoft, Google, and Amazon, are providing customized VA services to drive a variety of VA applications.

Although EVA has gained widespread attention in academia and industry in the past decade, there is no comprehensive and up-to-date survey on EVA. Among the most relevant surveys \cite{Survey1,Survey2,Survey3,Survey4,Survey5}, references \cite{Survey1,Survey2,Survey3,Survey4} concentrate on a broader topic: edge intelligence (EI), which is the convergence of edge computing and DL. However, due to their general focus on EI as a whole, insufficient details on EVA are provided; reference \cite{Survey5} briefly covers the concept of EVA, and introduces the basic components of a video analytics pipeline (VAP). However, the discussions on enabling techniques and EVA systems are limited to VA application cases in public safety. In addition, these surveys do not include the latest progress of EVA from 2020 to the present. Consequently, there is a need for a dedicated and comprehensive investigation into EVA, as it can shed light on the distinct challenges and opportunities within this area, and provide valuable insights for researchers and practitioners seeking to advance the state-of-the-art in EVA. Table \ref{surveycomparison} compares the scope of relevant surveys and this paper on EVA. Clearly, this paper differentiates itself from the earlier surveys in the following aspects:
\begin{enumerate}[wide]
    \item \textbf{Timeliness}: The field of EVA is fast-growing. This paper collects and surveys the most up-to-date research work in EVA as of Aug. 2023.
    \item \textbf{Nomenclature}: There are a number of related concepts in VA and EVA. Different (and sometimes confusing) terminologies have been used in existing literature. The survey elucidates key nomenclatures, which are crucial to the discussion of EVA applications and systems.
    \item \textbf{Application Categorization}: This paper introduces a novel categorization of VA applications based on multiple criteria. It offers a framework for researchers to better understand the different types of VA applications, their deployment scenarios and the associated unique challenges.
    \item \textbf{System Architectures}: This paper introduces different EVA system architectures. This topic was mentioned in the prior surveys, but the discussion was limited to the system architectures for DL, rather than EVA. In this paper, we provide a detailed discussion on EVA system architectures with an emphasis on their unique challenges beyond DL. Furthermore, inspired by \cite{killerapp}, this paper breaks down an EVA system into multiple basic components, and provides detailed descriptions and explanations of each component and how they work together. Doing so can fill a critical gap and offer insights for researchers and practitioners in developing and deploying EVA solutions.
    \item \textbf{Enabling Techniques}: While the prior surveys discussed techniques for optimizing edge computing systems, they primarily focused on general-purpose DL applications. In contrast, this paper provides comprehensive coverage of enabling techniques, especially those aiming to optimize the performance of EVA systems.
\end{enumerate}

The structure of this paper is shown in Fig. \ref{structure}. The remainder of this paper is organized as follows: Section II delineates the survey scope, paper selection criteria and results. Section III reviews the fundamentals of edge computing, including its definition, basic components and common computational hardware. Section IV offers an overview of VA, encompassing its definition, basic components and application categorization. Section V investigates EVA systems in depth, including the definition of EVA, system architectures, basic system components, performance indicators, and application architectures. Section VI discusses five enabling techniques for optimizing EVA systems: performance profiling, input filtering, hierarchical inference, configuration optimization, and workload placement and scheduling. Section VII introduces mainstream container orchestration frameworks and widely adopted datasets, providing essential resources and tools for researchers and practitioners in the development of EVA systems. Finally, Section VIII highlights challenges and potential future research directions.

\begin{table*}[htbp]
    \caption{Comparison between the prior surveys and this survey}
    \label{surveycomparison}
    \centering
    \begin{tabular}{|c|c|c|c|c|c|c|c|}
        \hline
        \textbf{Survey} & \textbf{Timeliness} & \textbf{Nomenclature} & \textbf{Application Categorization} & \textbf{System Architectures} & \textbf{Enabling Techniques}\\
        \hline
        \cite{Survey1} & Jan 30, 2020 & \ding{55} & \ding{55} & \ding{70} & \ding{70}\\
        \hline
        \cite{Survey2} & Jul 15, 2019 & \ding{55} & \ding{55} & \ding{70} & \ding{70}\\
        \hline
        \cite{Survey3} & Jun 12, 2019 & \ding{55} & \ding{55} & \ding{70} & \ding{70}\\
        \hline
        \cite{Survey4} & Apr 1, 2020 & \ding{55} & \ding{55} & \ding{55} & \ding{70}\\
        \hline
        \cite{Survey5} & Jul 30, 2019 & \ding{70} & \ding{55} & \ding{70} & \ding{70}\\
        \hline
        Ours & Aug 14, 2023 & \ding{51} & \ding{51} & \ding{51} & \ding{51}\\
        \hline
        \multicolumn{6}{l}{\ding{51} The topic is thoroughly covered; \ding{70} the topic is partially covered; \ding{55} the topic is not covered.}\\
    \end{tabular}
\end{table*}

\section{Paper Selection Methodology}
This section covers the survey scope, paper selection criteria and results.

\subsection{Survey Scope}
This paper delves into the core system aspects of EVA and emphasizes the critical importance of real-time performance in EVA systems. Exploring the synergistic interplay between VA and edge computing, it covers the design, optimization, and implementation of EVA systems that efficiently harness DL techniques. In addition, this paper dives into the enabling techniques that address the delicate balance between accuracy and latency, which is of paramount importance in real-time EVA applications. Through a comprehensive review of relevant research work and deployed systems, the paper aims to provide a deeper understanding of the factors influencing real-time performance in EVA systems. The basics of edge computing and CV are briefly discussed, since they serve as a foundation for understanding the subsequent materials. However, they are not the focus of this paper, as they have been extensively explored in existing surveys \cite{ODSurvey,motsurvey,mec,mec2}. Although privacy preservation \cite{privid}, training stage optimization \cite{Ekya,Mercury,mirsa,mistify,TTF,Federated}, and energy efficiency  \cite{AsyMo,GEMEL,Elf} are important to EVA, these topics fall outside the scope of this paper. The present investigation focuses on the system aspects and real-time performance of EVA systems, leaving the exploration of these additional challenges in future endeavours.

\subsection{Paper Selection Criteria}
To ensure comprehensive coverage of research papers on EVA in the scope identified, we first collect papers using the tool \textit{Collected Papers}, starting from several top-cited representative papers: \cite{Chameleon,videoEdge,VideoStorm,LAVEA,AWStream}. Then we apply keyword matching to search papers in the edge computing domain on Google Scholar. We focus on the most recent papers published from 2018 to the present, to reflect the most recent developments in this field. The keywords used are listed as follows: video analytics, vision analytics, video processing, video surveillance, stream processing, and stream analytics.

Furthermore, we manually browse the publications in top-tier conferences on relevant subject areas (e.g., USENIX ATC, SenSys, INFOCOM, SIGCOMM, MobiCom, SEC, NSDI, OSDI, etc.) and the corresponding workshops in the past five years to avoid the limitations of keyword matching.

\begin{figure}[b]
\centering
\includegraphics[scale=0.37]{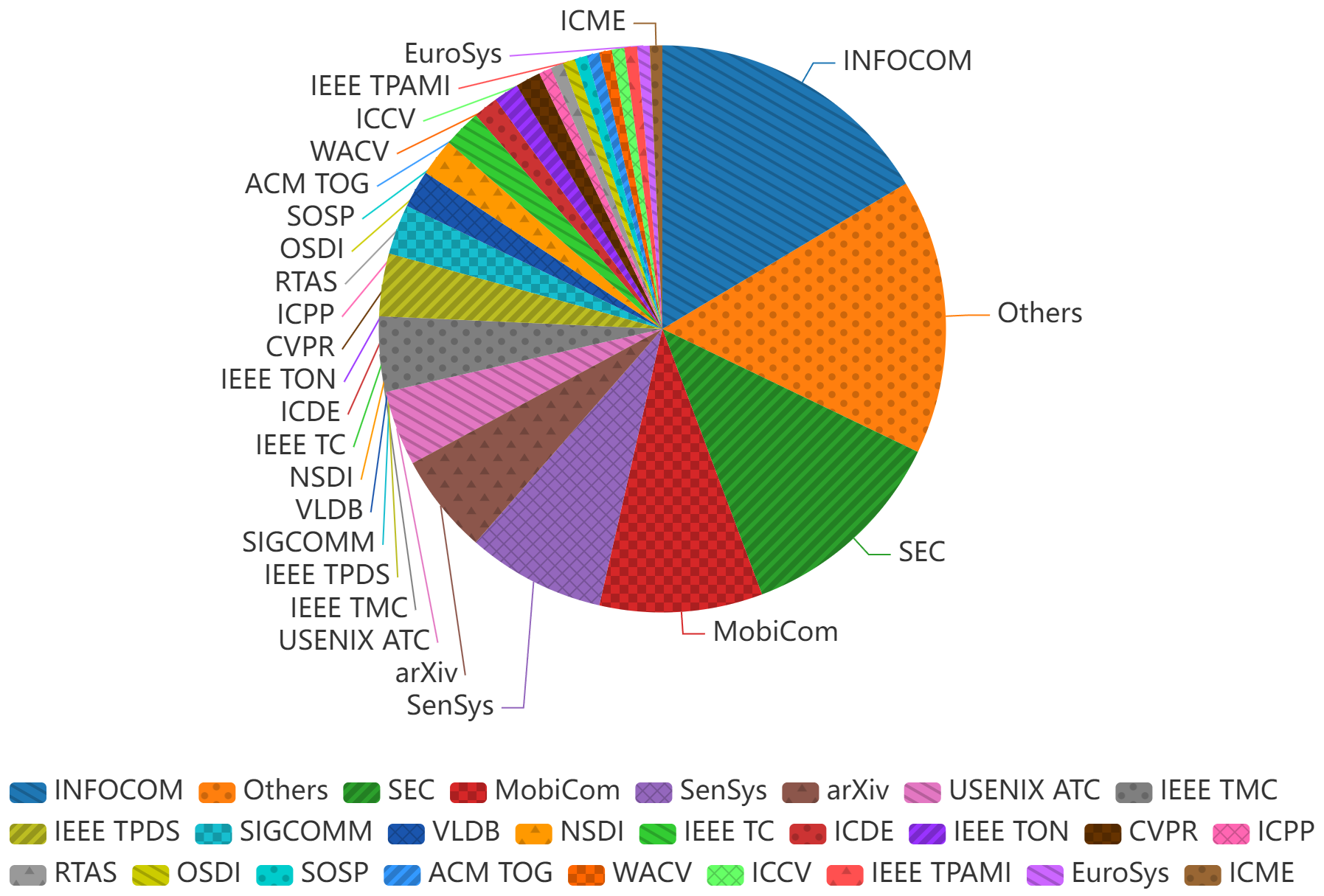}
\caption{Publication venues of the papers included in this survey.}
\label{venue distribution}
\end{figure}

\subsection{Paper Selection Results}
We start with 1446 papers collected from Google Scholar, DBLP, IEEE/ACM digital library and arXiv based on the aforementioned criteria. In the first round of screening, we eliminate 834 papers whose titles lack sufficient relevance. Next, we remove 389 papers that are not published in top-tier conferences or journals (i.e. ranked as A or A* according to CORE \cite{core}), except for those with over 100 citations. In the final round, we manually review the abstracts of the remaining papers to ensure that their topics match the survey scope. After three rounds of screening, we are left with 140 papers. All of them cover themes related to video analytics, computer vision, and edge computing topics. Fig. \ref{venue distribution} shows the distribution of papers published in different research venues. Here, “Others” includes papers published in venues not named. For unpublished papers in arXiv, we define them as the “arXiv” category.

According to Fig. \ref{venue distribution}, it can be observed that conference papers account for a larger proportion than journal papers, suggesting that conferences are preferred by most researchers to share their latest results in this domain. Among these conferences, INFOCOM accounts for the largest percentage, followed by SEC, MobiCom, and SenSys. It is important to note that SEC is an unranked conference based on CORE, but given its influence in the edge computing community and the context of this paper, we consider it an important venue on this topic.

\begin{figure}[tb]
\centering
\includegraphics[scale=0.31]{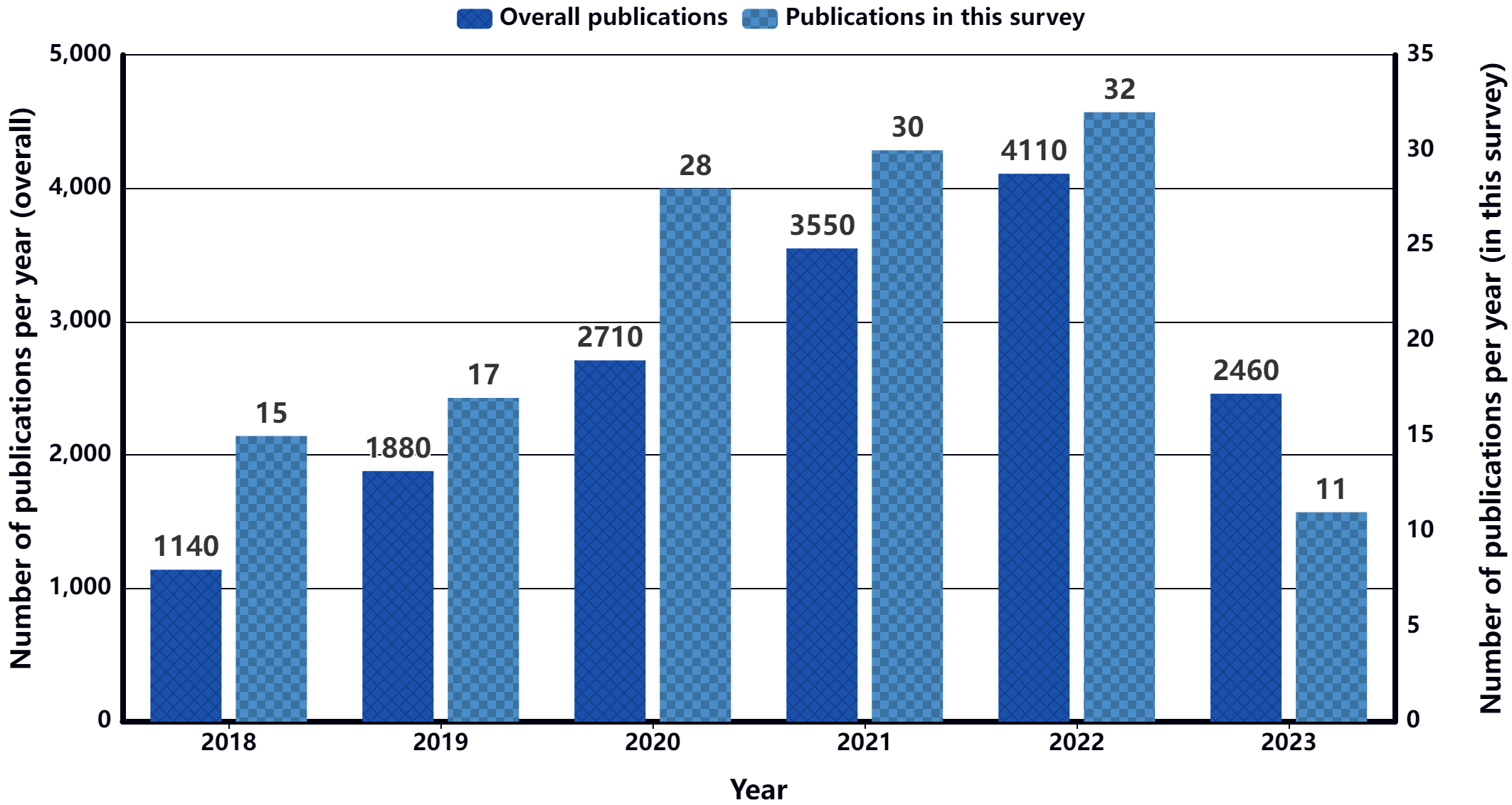}
\caption{Number of publications (overall and in this survey) on EVA over years. The overall numbers are from Google Scholar using the aforementioned keywords. Data for 2023 include publications up to August.}
\label{paper number}
\end{figure}

In addition, we conduct a study on the number of publications on EVA from 2018 to the present, by searching papers using the aforementioned keywords on Google Scholar. The results are shown in Fig. \ref{paper number}. It is evident that the EVA domain experienced a rapid increase of 261\% in the number of publications from 2018 to 2022. A noticeable drop of 40\% can be seen in 2023, as the latest papers might not have been included in Google Scholar as of the writing of this paper. The trend of papers included in this survey is consistent with the overall trend.

\begin{figure}[tb]
\centering
\includegraphics[scale=0.6]{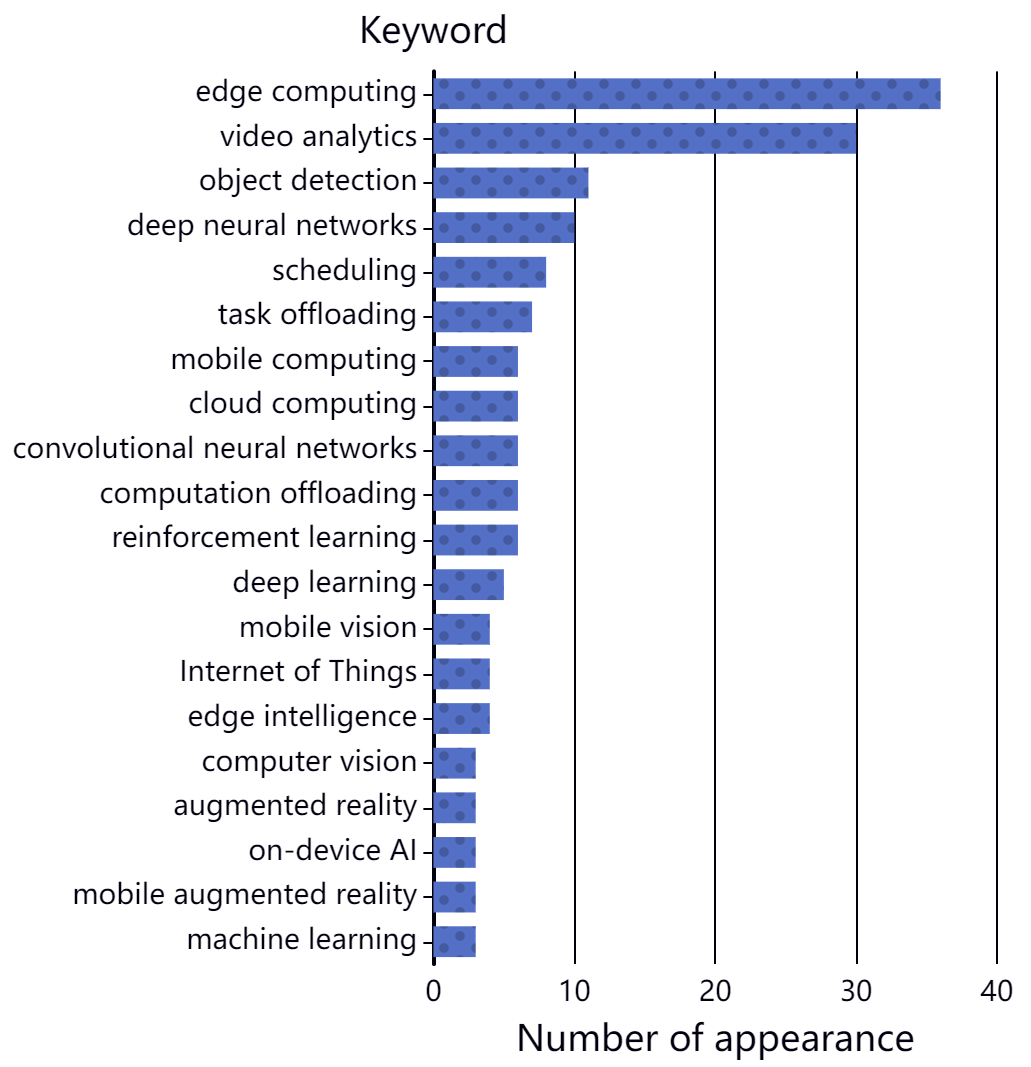}
\caption{Frequency of the keywords in the 140 surveyed papers. Only the top 20 keywords are listed.}
\label{word number}
\end{figure}

We also conduct a statistical analysis on the appearance frequency of the keywords of the 140 papers selected to understand the subject topics of these papers. According to Fig. \ref{word number}, unsurprisingly, we can see that “edge computing” has the highest frequency, followed by “video analytics”. These statistics are consistent with the keywords we use, and reflect the focus of this survey, namely, edge video analytics. Other keywords show related domains (e.g., “object detection”, “deep learning”, “computer vision”, “reinforcement learning”, “Internet of Things”, “augmented reality”, etc.), computing paradigms (e.g., “mobile computing”, “edge computing”, etc.), and key techniques (e.g., “scheduling”, “task offloading”, “computation offloading”, etc.).

\section{Fundamentals of Edge Computing}
Real-time data processing demand is driving the edge. IDC forecasts that by 2025, over 150 billion devices will be connected globally, with the majority generating real-time data \cite{IDC}. This translates to an average digital interaction every 18 seconds for every connected individual \cite{IDC}. Edge computing, as a promising solution, has attracted widespread attention \cite{cloudtoedge,cloudtoedge2}. According to \cite{IDC}, 68\% of financial services, 58\% of manufacturing, 48\% of healthcare, and 34\% of media and entertainment have utilized edge computing to manage real-time data. In this section, we introduce the fundamentals of edge computing.

\subsection{Definition of Edge Computing}
The earliest definition of edge computing was proposed in 2014 by Karim Arabi, the former vice president of research and development at Qualcomm as follows:

“\textit{All computing outside the cloud that happens at the edge of networks, and more specifically in applications where real-time processing of data is required.}”

Karim Arabi believed that cloud computing and edge computing should have distinct purposes: cloud computing focuses on processing big data while edge computing focuses on handling “instant data”, i.e., data generated by sensors or users in real-time.

The concept of “edge” sprang up even earlier, and it can be traced back to the 1990s when Akamai Technologies introduced content delivery networks (CDNs) to improve web performance by placing nodes at locations geographically closer to the end user \cite{edgeemergence}. These nodes cached web content (e.g., images and videos), accelerating visual content delivery and improving users' visual experience. In 2006, cloud computing emerged with the release of Amazon’s Elastic Compute Cloud (EC2) service, and clouds quickly became the most popular infrastructure for companies to run their businesses \cite{edgeemergence}. Since then, three paradigms, namely, \textit{Cloudlet Computing}, \textit{Fog Computing}, and \textit{Multi-Access (Mobile) Edge Computing} (MEC), symbolizing the early stage of edge computing began to emerge \cite{edgeemergence,edgeroadmap}:
\begin{enumerate}[wide]
    \item \textit{Cloudlet Computing}: The concept of cloudlet was originally proposed by Carnegie Mellon University in 2009 \cite{cloudletorigin}. A cloudlet is defined as a trusted, resource-rich computing infrastructure (e.g., a computer, a cluster of computers) that is well-connected to the Internet and can be accessed by nearby mobile devices via wireless local area networks (WLANs) \cite{cloudletorigin}. The key motivation of cloudlet is to improve the interactive performance of mobile applications, particularly those with rigorous requirements on end-to-end latency and jitter, such as language translation and facial recognition \cite{cloudletorigin}. To guarantee users' QoE, such applications require a response time on the order of milliseconds, which is highly difficult to achieve over WANs. Owing to the physical proximity and high communication speed of WLANs, cloudlets can provide highly responsive cloud services to mobile users and hence complement the three-tier hierarchy, i.e., end-cloudlet-cloud. The concept of micro data centers (MDCs) was first proposed by Microsoft in 2015 \cite{mdcorigin}, and the idea is similar to cloudlet \cite{Survey1, fogcloudlet, all}. Now, these two terms are almost equivalent.
    \item \textit{Fog Computing}: The term “fog computing” was first coined by Cisco in 2012, with the purpose of dealing with a huge number of IoT devices and massive data volumes for latency-sensitive applications \cite{fogorigin}. Extended from cloud computing, this new paradigm includes a three-tier architecture: end-fog-cloud, with the fog layer located between the cloud and end layers. The fog has cloud-like properties (e.g., sufficient resources, geographical distribution, heterogeneity, scalability, elasticity, etc.) and can provide the lowest possible service latency as it is aware of its logical location and closer to the “ground”, i.e., IoT devices \cite{fogorigin,fognist}.
    \item \textit{Multi-Access (Mobile) Edge Computing}: Mobile edge computing was initiated by the European Telecommunication Standards Institute (ESTI) to provide IT and cloud computing capabilities at the edge of the cellular network \cite{mecorigin,mec,mec2}. By placing MEC servers in the proximity of cellular base stations, MEC can improve the user experience by processing user requests at the network edge with lower latency, location awareness and network context-related services (e.g., local points-of-interest, businesses and events) \cite{mecorigin}, as well as alleviate the load on the core network. Currently, the terminology “MEC” has been extended by ESTI from mobile edge computing to multi-access edge computing by accommodating wireless communication technologies (e.g., Wi-Fi) \cite{Survey1}.
\end{enumerate}

\begin{figure}[tb]
\centering
\includegraphics[width=0.47\textwidth]{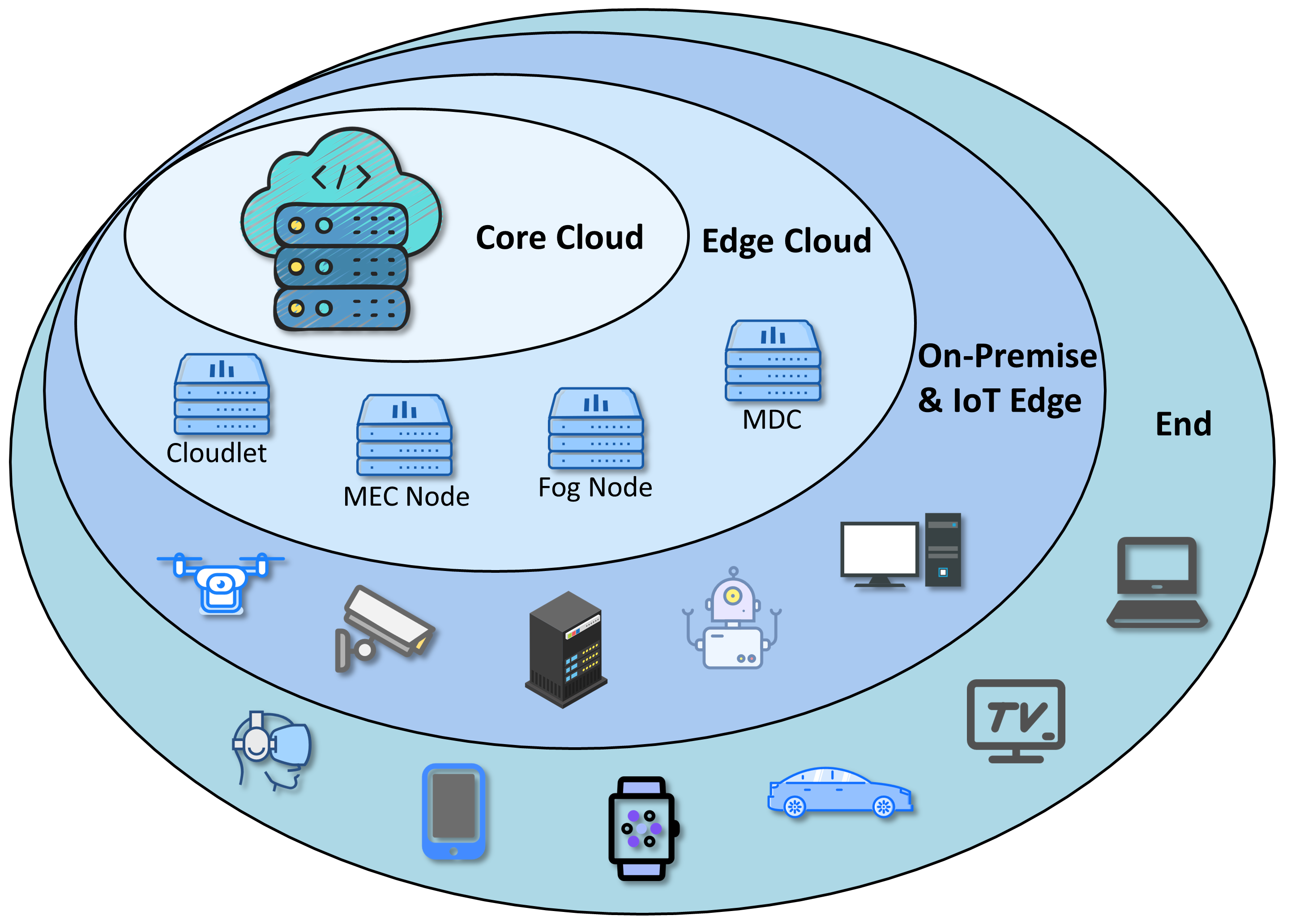}
\caption{Illustration of Edge Computing.}
\label{edge computing}
\end{figure}

\subsection{Basic Components in Edge Computing Paradigm}
Due to the rapid development of edge computing, the community has not yet reached a consensus on standard definitions, architectures, and protocols of edge computing \cite{Survey1}. The boundaries between cloudlet, MDC, fog, MEC and edge are blurred \cite{edgeemergence}. Reference \cite{Survey1} simply uses the term “edge” to represent all of them. We follow this approach and divide the edge computing paradigm into three layers, as illustrated in Fig. \ref{edge computing}:
\begin{enumerate}[wide]
    \item \textit{End}: In the context of edge computing, an end device is a source device in a networked system, which generates data by interacting with users (e.g., mobile phones, laptops, etc.) or sensing environments (e.g., cameras, sensors, etc.). Considering that modern end devices usually carry computing capacities, they are able to process workloads locally without offloading them to remote servers. For example, Tesla enables autonomous driving by leveraging in-vehicle computers to analyze data from sensors and cameras surrounding the car. AWS DeepLens, which is the world's first DL-enabled video camera, can host DL applications such as face detection and activity recognition on cameras. Therefore, modern end devices contain both the end part, which generates data and the edge part, which processes data from the end part.
    \item \textit{Edge}: An edge device can be any processing unit that is in close proximity to end devices and capable of processing data generated from them. Communication between end and edge devices utilizes protocols like Wi-Fi, Bluetooth, Zigbee, 4G/5G, or Ethernet, depending on application requirements, such as data throughput, latency, and range. Based on scale, we divide edge into three categories, i.e., IoT edge, on-premise edge and edge cloud. Notably, some of the IoT edge and on-premise edge overlap with the end, since they could generate and process data locally (e.g., smart cameras, desktops). The descriptions of these three categories are presented as follows:
    \begin{itemize}
        \item \textit{IoT Edge}: IoT Edge covers almost any device that 1) can communicate with remote entities via networks like local area networks (LANs) or WANs, and 2) has limited computation and storage capabilities to process data locally. As shown in Fig. \ref{edge computing}, representative examples of IoT edge include drones, smart cameras, robots, etc.
        \item \textit{On-Premise Edge}: On-premise edge refers to dedicated computational, network, and storage infrastructure owned and managed by specific organizations, such as hospitals, retail stores, shopping malls, university campuses, corporate buildings, or manufacturing plants. The infrastructure can provide various services to authenticated users (e.g., employees, students, registered customers), including data processing, data storage, network routing, privacy and security preservation, application hosting, etc. On-premise edge devices can take many forms, ranging from a single computer to a data center housing multiple racks of servers.
        \item \textit{Edge Cloud}: While cloud computing excels in resource-intensive data processing and workloads like machine learning model training, its latency can be problematic for some latency-sensitive workloads as data must travel all the way to a remote data center and back again. To allow real-time service responses, cloud service providers are extending their services from the network core to the network edge by placing vast and highly-distributed edge clouds closer to local users, with software to deliver services in a way that is similar to using public cloud services. One use of the edge cloud is the delivery of visual content. Streaming services, cloud gaming, augmented reality and other visual workloads are growing dramatically. Computing services hosted in edge clouds can help cope with these visual workloads at locations closer to the customers, resulting in reduced latency and enhanced customer experiences. Therefore, the edge cloud can be viewed as an extension and complement to the core cloud \cite{all,edgecloud}. 
    \end{itemize}
    
        The three types of edges have their own advantages and disadvantages. IoT edges can deliver the fastest responses since they are co-located with the data source, but constrained by their limited resources, they cannot handle heavy workloads. On-premise edges have more resources than IoT edges, and can produce rapid responses due to their proximity to the data source, but they require separate setup and configuration for each location and device, along with staff to manage them. Provisioned by major cloud service providers, edge clouds typically have sufficient resources, and require no installation and maintenance costs, but because they are farther away and need to be accessed by WANs, the latency can be higher than IoT edges and on-premise edges.

    \item \textit{Cloud}: Cloud, or core cloud, refers to remote large-scale data centers with powerful computing capabilities and massive storage space that can provision a large number of scalable and elastic virtual machines (VMs). Based on the definition of the National Institute of Standards and Technology (NIST) \cite{cloudnist}, a core cloud has five essential characteristics: 1) on-demand self-service, 2) broad network access, 3) resource pooling, 4) rapid elasticity and 5) measured service. A core cloud is a controller and orchestrator of all distributed edge clouds, while the edge clouds inherit all the characteristics of the core cloud.
\end{enumerate}
    
\subsection{Computational Hardware}
In this section, we discuss enabling hardware for running artificial intelligence (AI) workloads at the edge, encompassing general-purpose processing units such as central processing units (CPUs) and graphics processing units (GPUs), as well as customized hardware for AI workloads like field programmable gate arrays (FPGAs) and application-specific integrated circuits (ASICs).
\begin{enumerate}[wide]
    \item \textit{Central Processing Unit}: CPUs are standard and general-purpose processors used in many devices, playing an important role in VA, especially when real-time processing is not a primary concern. While CPUs may not be as efficient as GPUs for highly parallelized tasks, especially large-scale matrix operations, they are instrumental in handling tasks like video decoding, basic motion detection, and simple VA algorithms. For instance, in surveillance systems where only motion detection is required, a CPU might suffice. Moreover, the energy efficiency of CPUs (especially ARM processors) makes them suitable for battery-powered edge devices, such as mobile phones, which are increasingly being used for lightweight VA tasks, like face recognition or basic augmented reality (AR) applications.
    \item \textit{Graphics Processing Unit}: GPUs, due to their parallel processing capabilities, have become the “engines” of many advanced VA applications. Their highly parallel processor architecture is particularly suited for computation-intensive DL tasks like object detection, scene segmentation, and video recognition. For instance, large-scale real-time tracking and counting for road users (e.g., pedestrians, bicycles, vehicles, etc.) in traffic monitoring systems heavily rely on GPUs \cite{rocket,killerapp}. NVIDIA's Compute Unified Device Architecture (CUDA) platform has further cemented the GPU's position in VA by enabling developers to optimize their algorithms for GPU architectures, ensuring faster video processing and real-time insights.
    \item \textit{Field Programmable Gate Array}: FPGAs, given their reconfigurable nature, are emerging as powerful tools in VA. They can be tailored to handle diverse situations with low added development costs. For instance, in traffic monitoring systems, FPGAs can be programmed to detect and track vehicles, adjusting their logic to cater to different lighting conditions or vehicle types. Their ability to host multiple functions in parallel also means that a single FPGA can handle video decoding, object detection, and data transmission simultaneously, making them ideal for complex VA setups \cite{TAPU,Gemini}. Recently, FPGAs have been increasingly used as AI workload accelerators. A representative example is Microsoft's “Project Catapult” \cite{catapult}, where FPGAs are employed to provide ultra-fast inference at the edge.
    \item \textit{Application-Specific Integrated Circuit}: ASICs, being highly specialized, are the pinnacle of efficiency when it comes to specific VA tasks. Google's Edge Tensor Processing Unit (TPU) \cite{TPU,edgetpu}, for instance, is optimized for running lightweight vision models on edge devices, making it a good candidate for applications like real-time object detection. Similarly, Intel's Movidius Vision Processing Unit (VPU) is designed for high-performance video processing, which finds applications in drones or robots that rely on live video feeds to navigate or interact with the environment. Several other companies, including IBM, HUAWEI, Cambricon Technologies, and Horizon Robotics, are actively engaged in the development of their proprietary AI ASICs. For instance, IBM has introduced TrueNorth (Neural Processing Unit, NPU) \cite{truenorth}, while Huawei is advancing with its Ascend AI chips (NPU), Cambricon is producing chips with NPU capabilities, and Horizon Robotics is pioneering Sunrise and Journey (Brain Processing Unit, BPU)
\end{enumerate}

Overall, each type of computational hardware is suited for a particular kind of workload, and using them together in heterogeneous computing applications provides all of the functionality that complex use cases require. When combined, they can also balance workloads, boost different AI inference performances, and build the most effective and efficient configurations \cite{chipselection}.

\section{Overview of Video Analytics}
In this section, we begin by clarifying the definition of VA, since the terminology has different meanings in different contexts. We then introduce the basic components of a VAP, followed by a taxonomy for VA applications.

\subsection{Definition of Video Analytics}
Video analytics, also known as video content analysis, refers to the process of automatically extracting valuable information and insights from video data using techniques such as CV and DL. It involves recognizing patterns, detecting objects and tracking movements in order to analyze, interpret, and understand video content and make data-driven decisions. In real life, VA is applied in diverse domains, such as security and surveillance for camera feed monitoring and intrusion detection, traffic management for traffic flow optimization and incident detection, retail analytics for customer behaviour monitoring and store layout management, manufacturing and industrial applications for product defect identification and worker safety, sports analytics for player movement and team strategy assessment, and healthcare applications for patient behaviour monitoring, fall detection, and medical equipment tracking.

\begin{figure}[htbp]
\centering
\subfloat[]{
\includegraphics[width=0.47\textwidth]{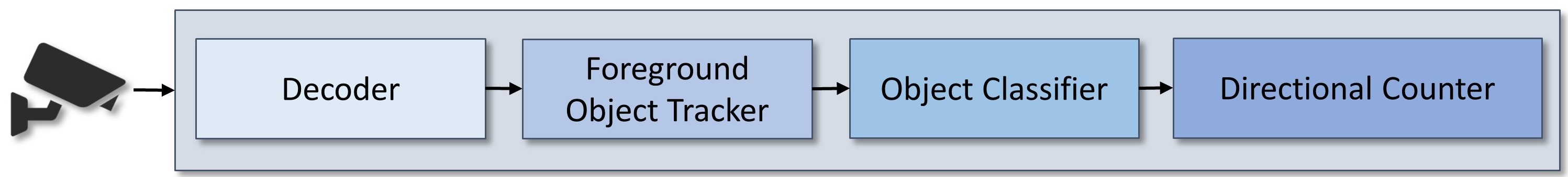}
\label{VAP example 1}}
\hfil
\subfloat[]{
\includegraphics[width=0.47\textwidth]{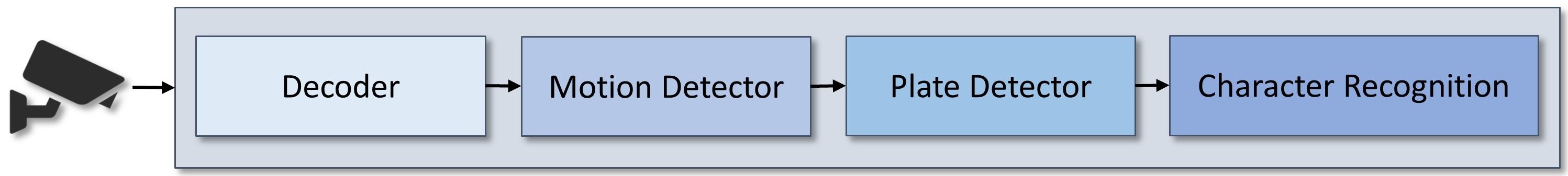}
\label{VAP example 2}}
\caption{Two examples of VAPs. (a) Pipeline of a vehicle counting application. (b) Pipeline of a license plate recognition application.}
\end{figure}

A VAP refers to a series of sequential steps or stages through which the data passes to be transformed, analyzed, and processed in a VA application. The pipeline represents the overall flow and organization of the various operations and algorithms applied to the data, from the initial input to the final output. Each stage in the pipeline typically focuses on a specific task or function, and the output from one stage becomes the input for the next stage, allowing for a modular and structured approach to processing the data. Typically, a VAP is composed of multiple video processing modules, which can vary across applications. For instance, the pipeline of a vehicle counting application (Fig. \ref{VAP example 1}) consists of a video decoder, followed by a foreground object tracker, an object classifier and a directional counter. In contrast, in a license recognition application, the pipeline (Fig. \ref{VAP example 2}) may consist of a video decoder, a motion detector, a plate detector and a character recognition module. Therefore, the components of a VAP are \textit{application-dependent}.

\subsection{Basic Components of Video Analytics Pipelines}
Computer vision is foundational to VA, as VAPs are constructed on the principles and techniques of CV. Despite diverse application requirements, several CV techniques are ubiquitously employed across VAPs. In the following discussion, we provide a concise overview of these essential techniques. For a more in-depth exploration, readers are encouraged to consult the dedicated literature \cite{ODSurvey,ODSurvey2,motsurvey,motsurvey2,dlsurvey,segsurvey,ocsurvey,dlsurvey2,cnnsurvey}.

\begin{enumerate}[wide]
\item \textit{Video Encoding and Decoding}: Encoding is a crucial component in video streaming. It significantly reduces communication overhead by compressing multiple consecutive frames into a compact segment or chunk. As reported in \cite{OsmoticGate}, encoding can accelerate video streaming by 10$\times$, compared to streaming the raw frames one by one. The prevalent video encoding formats include advanced video coding (H.264), high-efficiency video coding (H.265), versatile video coding (H.266), etc. Encoded videos are often streamed based on live streaming protocols, such as real-time messaging protocol (RTMP), real-time transport protocol (RTP), real-time streaming protocol (RTSP), secure reliable video transport (SRT), web real-time communication (WebRTC), HTTP live streaming (HLS), and dynamic adaptive streaming over HTTP (MPEG-DASH). To reconstruct the original frames, video decoding is employed at the edge or cloud to decompress the encoded videos.

\item \textit{Pre-processing}: After decoding, videos are pre-processed before being subject to further analytics. The pre-processing operations include image resizing \cite{EdgeVision}, cropping \cite{DeepStream2}, super-resolution \cite{EADV}, denoising \cite{elasticedge,EADV,imgpreprocessing}, deblurring \cite{imgpreprocessing}, dehazing \cite{dehaze}, and deraining \cite{imgpreprocessing}. In general, these operations aim to improve the view quality and can therefore benefit the subsequent procedures. In real-world applications, a video may contain multiple sources of noise. For instance, traffic videos captured at night may suffer from poor illumination and motion blur. OpenCV \cite{opencv}, a well-known library in image and video analysis, implements a wide range of pre-processing algorithms for image denoising, resizing, rotating, padding, normalization, color space conversions, morphological operations, background subtraction for video motion detection, etc. In addition to conventional image processing techniques, generative adversarial networks (GANs) have been utilized in image enhancement \cite{Turbo}. Another way is to directly remove those frames with high distortions, because such frames are not likely to contribute to downstream tasks \cite{AQuA}. As reported in \cite{SMOL}, pre-processing can be a bottleneck in many VA systems on modern hardware. Turbo \cite{Turbo} optimizes this process by employing a hierarchical model to selectively enhance incoming frames at different levels based on latency and resource budget. CV-CUDA \cite{CV-CUDA}, a recent library still under development, tries to leverage GPUs to accelerate pre-processing operations.

\item \textit{Object Classification}: Object classification, also known as object recognition or object identification, maps an object into one of a finite set of classes. Early object classification is primarily based on handcrafted features and shallow machine learning (ML) models. Methods such as Scale-Invariant Feature Transform (SIFT) and Histogram of Oriented Gradients (HOG) are used to extract features from images, which are then fed into classifiers like Support Vector Machines (SVM) or k-Nearest Neighbors (k-NN) \cite{classification_HOG,classification_SIFT}. These methods, while effective for certain tasks, often struggle with variations in lighting, pose, and scale. With the advent of DL, the paradigm shifted towards end-to-end learning. Convolutional neural network (CNN)-based classifiers, such as ResNet \cite{ResNet} and MobileNet \cite{MobileNets}, have become dominating approaches, leveraging large labeled datasets to predict the class of target objects with remarkable accuracy, surpassing traditional ML-based approaches.

\item \textit{Object Detection}: Object detection involves locating and recognizing objects in frames. Traditional object detection methods first identify regions of interest (RoIs), using image processing methods like background subtraction \cite{od_background}, frame differencing \cite{od_fd}, and optical flow \cite{od_of,od_of2}. Once these regions are detected, they are classified using an object classifier. Nowadays, object detection algorithms typically leverage deep neural networks (DNNs) to achieve high accuracy and can be classified into two categories: two-stage and one-stage \cite{ODSurvey}.
\begin{itemize}
    \item \textit{Two-stage}: A two-stage object detector first extracts RoIs, and then makes a separate prediction for each of these regions. Faster region-based convolutional neural network (RCNN) \cite{FasterRCNN} represents a classical two-stage detector. It employs a region proposal network (RPN) to generate region proposals and performs classification on these regions separately. 
    \item \textit{One-stage}: A one-stage object detector, in contrast, simply applies a single DNN model for both object localization and recognition. The two tasks are cast as a unified regression problem. The most widely-known one-stage detectors include the YOLO family \cite{Yolo1,Yolo2,Yolo3} and the SSD family \cite{SSD}. In general, one-stage detectors are much faster but less accurate than two-stage ones.
\end{itemize}

The types of objects of interest can vary across applications. For example, in traffic monitoring applications, we are concerned with road users, such as pedestrians and vehicles \cite{VeTrac,Chameleon2}, while in retail surveillance applications \cite{VAretail,MCMOT1}, customers and merchandise need to be detected. Other interesting objects in the VA domain include plants \cite{visage}, animals \cite{animal}, faces \cite{EagleEye}, postures \cite{vaposture}, activities \cite{Caesar}, flames \cite{vafire}, smoke \cite{vasmoke}, etc. Therefore, it is advantageous to tailor generic object detectors and classifiers to target deployment scenes.

\item \textit{Object Tracking}: Object tracking is the process of locating objects and estimating their trajectories from a video sequence. Conventional methods follow the tracking-by-detection paradigm \cite{Tracking}, performing multi-object tracking (MOT) sequentially using two separate models \cite{sort,deepsort}. A detector first detects bounding boxes of objects in each frame, after which a re-identification (re-ID) model extracts visual features from each bounding box and links the objects based on these features and motion cues. Such two-stage methods, while effective, are computationally intensive, especially when the scene is crowded. Recent advancements in multi-task learning have led to joint models where detection and re-ID tasks share a common backbone, significantly reducing inference time. These integrated models are often termed one-shot trackers \cite{mots,JDE,centertrack,fairmot}.

\item \textit{Image Segmentation}: Image segmentation involves partitioning an image into multiple segments, each representing a distinct object or region. It simplifies and changes the image representation into something more meaningful and easier to analyze. Among the various types of image segmentation, semantic segmentation assigns each pixel to a specific class, while instance segmentation classifies each pixel and differentiates between distinct instances of the same object class \cite{segsurvey}. Panoptic segmentation, on the other hand, unifies the tasks of semantic and instance segmentation, producing coherent labelling of all pixels, considering both regions and objects \cite{segsurvey}. These techniques can be integrated with other CV tasks, such as object detection and tracking, to achieve a more comprehensive understanding of scenes.

The aforementioned components need to be customized in practice based on the requirements of target applications. Users can specify additional attributes or constraints for relevant objects, e.g., counting the customers within a time period, detecting hazardous behaviours in public space, querying specific objects (e.g., a man dressed in blue), etc. The final results will be aggregated to answer user queries based on custom rules (so-called \textit{business logic}).
\end{enumerate}

\begin{table*}[tb]
    \caption{Various Edge Video Analytics Applications in Different Domains}
    \label{EVAapps}
    \centering
    \begin{tabular}{|c|c|c|c|c|c|c|c|c|}
        \hline
        \textbf{Application} & \textbf{Domain} & \textbf{Criterion 1} & \textbf{Criterion 2} & \textbf{Criterion 3} & \textbf{Criterion 4} & \textbf{Accuracy} & \textbf{Latency}\\
        \hline
        Traffic Analysis & Transportation & Retrospective & N/A & N/A & N/A & Medium & Low\\
        \hline
        Autonomous Driving & Transportation & Live & Multi-Camera & Continuous & Non-Stationary & High & High\\
        \hline
         Parking Lot Surveillance & Transportation & Live & Single-Camera & Continuous & Stationary & Medium & Medium\\
        \hline
        Traffic Incident Detection & Transportation & Live & Multi-Camera & Event-Driven & Stationary & High & High\\
        \hline
        Customer Behaviour Monitoring & Retail & Live & Multi-Camera & Continuous & Stationary & High & Medium\\
        \hline
        Product Placement Optimization & Retail & Retrospective & N/A & N/A & N/A & Medium & Low\\
        \hline
        Fall Detection & Healthcare & Live & Single-Camera & Continuous & Stationary & High & High\\
        \hline
        Robotic Surgery & Healthcare & Live & Single-Camera & Continuous & Non-Stationary & High & High\\
        \hline
        Front Door Surveillance & Home Automation & Live & Single-Camera & Event-Driven & Stationary & Medium & Medium\\
        \hline
        Pet Activity Monitoring & Home Automation & Live & Multi-Camera & Event-Driven & Stationary & Medium & Medium\\
        \hline
    \end{tabular}
\end{table*}

\subsection{Categorization of Video Analytics Applications}
Video analytics applications can be divided into multiple categories based on different criteria, e.g., the latency requirements, the number of cameras, the processing frequency and if the camera is moving. In this section, we present a taxonomy for VA applications, by classifying them based on four criteria: live or retrospective, single-camera or multi-camera, continuous or event-driven, and stationary or non-stationary. Table \ref{EVAapps} summarizes the categorization of ten representative EVA applications along with their accuracy and latency requirements.
\begin{enumerate}[wide]
    \item \textit{Live or Retrospective}: Live video analytics (LVA), also known as real-time VA, refers to the process of analyzing video data as it is captured, enabling immediate decision-making and action based on the information derived from the video stream. This approach is particularly beneficial in scenarios where timely responses are crucial. LVA requires all processes to be finished within the deadline (i.e., a relatively short time period) \cite{RT-mDL,deeprt}. For instance, if the video is streamed at 30 frames per second (FPS), to enable real-time responses, each frame should be processed within 33 milliseconds (i.e., the deadline); otherwise, the deadline will be missed, and the next frame will be stored in the buffer, waiting to be processed. In latency-sensitive applications, such as traffic accident detection and robotic surgery, citizens' safety could be compromised if the deadline is missed. On the other hand, retrospective video analytics (RVA), also known as offline or historical VA, involves the analysis of recorded video data, typically stored in a database or archive \cite{NoScope,TAHOMA,Blazeit,ExSample,THIA,Spatula,TASTI,Focus,DIVA,Boggart,cova}. This approach allows for a more in-depth examination of historical video data, enabling the identification of trends, patterns, and relationships that may not be apparent during real-time analysis. RVA is particularly useful for applications such as post-event investigations, long-term behavioural studies, and data-driven decision-making. For example, by analyzing archived video data, retailers can understand trends in foot traffic, dwell time in specific areas, and interactions with products. These insights can lead to better decisions in store layout, inventory management and promotional strategies.
    \item \textit{Single-Camera or Multi-Camera}: Single-camera VA involves the analysis of video data captured by a single camera, focusing on the detection, recognition, and tracking of objects or events within its field of view (FoV). This approach is typically employed when the area of interest can be sufficiently covered by one camera, and the primary objective is to gather insights or detect events within that specific area. For example, a camera is installed in a small parking lot to provide real-time information on parking availability to drivers and facility managers. Multi-camera VA, in contrast, involves the simultaneous analysis of video data from multiple cameras, which can be deployed in the same location with overlapping FoVs \cite{CrossRoI,Vigil}, or distributed in different locations without overlapping FoVs \cite{Chameleon,Caesar}. This approach offers a more comprehensive understanding of the environment, enabling the tracking of objects or events across different camera views and providing a holistic perspective of the area under surveillance \cite{MCMOT1}. By integrating and correlating the information from multiple cameras, multi-camera VA can deliver more accurate and reliable insights, improving the overall performance \cite{CrossRoI,Chameleon,vitrack,Anveshak}. Imagine in a large-scale surveillance application of a shopping mall, targets can frequently be missed by a single camera due to crowded scenes and obstacles. With multiple cameras, the lost targets are more likely to be captured by other cameras since every camera has a different perspective.
    \item \textit{Continuous or Event-Driven}: Continuous VA is the process of analyzing video feeds on a continuous basis without any specific trigger or event. This approach is commonly used in applications such as surveillance, where it is important to monitor a scene or environment continuously and identify potential threats or anomalies as they occur. For example, in a traffic monitoring application, cameras placed at key intersections or along highways can continuously monitor traffic patterns, detecting potential congestion, accidents, or other issues. The data can then be used by traffic management systems to optimize traffic flow and reduce travel time for drivers. Event-driven VA, on the other hand, processes video feeds in response to specific events or triggers. This approach involves using sensors or other devices to detect events such as motion, sound, or temperature changes, and then analyzing the video data associated with the event to extract insights or identify patterns. It is commonly used in applications such as smart home automation, where it is important to monitor specific events or behaviours and take appropriate actions in real time. For instance, when the doorbell camera detects motion or someone rings the doorbell, the VA system can be triggered to analyze the video feeds and identify who is at the door. This could be done using facial recognition technology. The system can then send a notification to the homeowner's smartphone, allowing them to see who is at the door and communicate with the visitor through a two-way audio system.
    \item \textit{Stationary or Non-Stationary}: Stationary cameras such as surveillance cameras and traffic cameras are fixed in one location, capturing a consistent FoV. They usually have optimized configurations (brightness, contrast, color-saturation, sharpness, etc.), mounting height, and shot angle based on the target environment. But for non-stationary cameras such as body-worn, drone and dash cameras, locations, surrounding objects, environments and shot angles constantly change, which can make it more difficult to identify objects or analyze the environment. Consequently, pre-processing techniques, e.g., rotation correction, background subtraction, noise reduction, and motion blur elimination, are often utilized to guarantee the quality of video frames before further processing. In addition, it is more challenging for non-stationary cameras to offload workloads due to the dynamic nature of their connectivity and location. They may experience intermittent or fluctuating network connectivity due to factors like distance, signal strength, or obstructions. For real-time processing, computations are mainly performed locally on cameras or co-located processing units (e.g., in-vehicle computers). To this end, lightweight pipeline designs and model-level optimization techniques targeting energy, memory, and storage efficiency (e.g., model compression \cite{amvp,NestDnn,mobisr}, scaling \cite{RT-mDL,legodnn}, merging \cite{GEMEL}, etc.) should be considered.
\end{enumerate}

\section{Edge Video Analytics System}
Large-scale real-time VA is becoming the “killer app” for edge computing \cite{killerapp}. First, the latency requirements for video processing can be stringent when the output of the analytics is used to interact with humans (e.g., virtual or augmented reality) or to actuate some mission-critical systems (e.g., security alerts and traffic lights). Second, transmitting high-definition videos requires substantial bandwidth (e.g., 5 Mbps or even 25 Mbps for 4K video) \cite{killerapp}. Streaming a large number of video feeds directly to the cloud is not always feasible since the available uplink bandwidth is often limited when the cameras are connected wirelessly, e.g., via cellular data networks inside vehicles. Finally, model inference is compute-intensive and naively performing inferences on cameras is inefficient as their computing capabilities are limited. We are likely to be stuck in the dilemma where using heavy DNNs can guarantee the accuracy requirements but miss the latency requirements, and vice versa for lightweight DNNs. Because of high data volumes, resource demands, and latency requirements, cameras are the most challenging “things” in the Internet of Things. Tapping into the convergence of edge computing and video analytics presents potential system challenges. In this section, we first clarify the definition of EVA. Then, we go through system aspects, including system architectures, basic system components, performance indicators, and application architectures.

\subsection{Definition of Edge Video Analytics}
\begin{figure}[tb]
\centering
\includegraphics[width=0.45\textwidth]{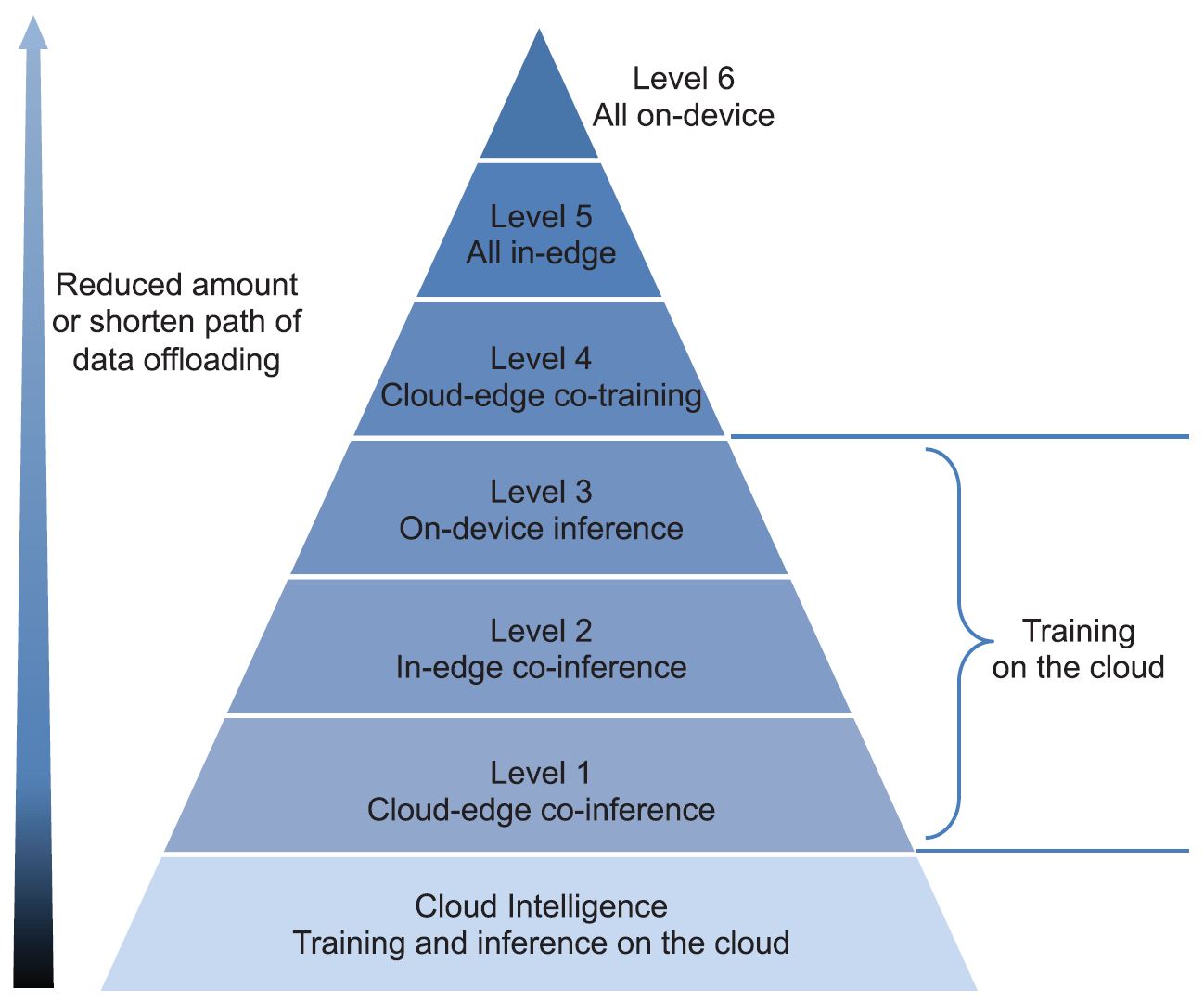}
\caption{Six-level rating for edge intelligence \cite{Survey3}.}
\label{EI rating}
\end{figure}

Currently, there is not yet a consensus on the exact definition of EVA. In \cite{Survey3},  Zhou et al. define edge intelligence, as the convergence between edge computing and artificial intelligence that fully exploits the available data and resources across the hierarchy of end, edge and cloud devices to optimize the overall performance of DL applications. Specifically, the authors rate EI into six levels, as shown in Fig. \ref{EI rating}. Since the scope of this paper does not cover model training, we only concentrate on level 1 to level 3, where training is done in the cloud.

Since EVA is a subset of EI, we define EVA similarly to EI, namely, as a paradigm that leverages the hierarchy of end, edge and cloud devices to maximize the performance of VA applications. By intelligently exploiting the most appropriate hierarchy based on specific application scenarios, EVA aims to improve the accuracy, efficiency, scalability, and responsiveness of VA systems. As a distinct subject within the broader domain of VA, EVA specifically focuses on addressing the challenges associated with real-time video processing, data transmission, and system optimization, thus enabling more effective and adaptable solutions to a variety of VA applications.

The Internet of Video Things (IoVT), as an emerging concept, is closely related to EVA. IoVT is a subdomain of IoT which refers to the integration of visual data within the framework of the IoT ecosystem \cite{iovt1,iovt2}. It encompasses an interconnected network of large-scale visual sensors, such as cameras (e.g., visible light, infrared, and thermal cameras) and lidar sensors that capture, process, and transmit data over the Internet. Recent work in EVA primarily focuses on improving the system and application performance, such as designing efficient VAPs (e.g., remove redundant computations \cite{ApproxNet}), increasing resource utilization (e.g., load balancing \cite{videoEdge}), and optimizing execution efficiency (e.g., on-device workload scheduling \cite{RT-mDL}). IoVT, on the other hand, deals with the entire ecosystem of visual sensors, analytics, and communication networks (including BSs, MEC servers, etc.). For example, \cite{SCA,MDUcast} study efficient large-scale video streaming based on 5G technologies, while \cite{JGO,JFEC,iovt_parking,iovt-SDN,iovt_counting} investigate IoVT analytics, a combination of IoVT and VA. While IoVT faces similar challenges to EVA, optimizing communication efficiency for multi-user VA systems is the primary focus of the limited existing work on this topic. This is often addressed using technologies such as Time Division Multiple Access (TDMA), which allocates distinct time slots for different users, Orthogonal Frequency Division Multiple Access (OFDMA) that enhances spectrum utilization through orthogonal subchannels, and Non-Orthogonal Multiple Access (NOMA), promoting spectrum sharing and variable power allocation to mitigate doubly near-far effects \cite{JGO,mec_survey}.

\subsection{Architectures of Edge Video Analytics Systems}
An EVA system is a specialized solution designed to hold one or multiple VA applications. It aims to optimize video processing pipelines, resource utilization, execution efficiency, data transmission and storage. An EVA system should be designed to function across a wide spectrum of deployments, and can vary from distributed analytics in a hybrid edge and cloud (level 1), to analytics across on-premise edges and IoT edges (level 2), to even hosting all analytics on IoT edges (level 3). In this section, we introduce four EVA system architectures, i.e., 1) pure-IoT edge, 2) IoT edge + on-premises edge, 3) IoT edge + edge cloud, and 4) IoT edge + on-premise edge + edge cloud, as illustrated in Fig. \ref{EVA arch}. The characteristics of each architecture are described as follows:

\begin{enumerate}[wide]
\item \textit{Pure-IoT Edge}: This architecture only includes IoT edges, namely, IoT devices such as mobile phones, surveillance cameras, smart vehicles, etc. As stated earlier, an IoT edge can be divided into an end part and an edge part. The end component collects data from the surrounding world, whereas the edge component performs computations locally on devices. For example, the latest smartphones can locally perform tasks such as real-time face recognition \cite{mobile_face} and rudimentary augmented reality. However, since IoT devices are usually resource-poor, they can only support light computation in real-time. To further enhance the accuracy and accelerate the computation, multiple IoT edge devices can form a mini-cluster and collaborate with each other to share information and resources \cite{camcluster,VisionPaper}. In the applications of vehicle-to-vehicle (V2V) communication systems, smart vehicles equipped with visual sensors can form ad-hoc networks to offload workloads and share information about traffic conditions, potential hazards, or optimal routes, enhancing the network resiliency to disruptions and improving safety and efficiency \cite{v2v,v2v_survey}.

\begin{figure}[tb]
\centering
\includegraphics[width=0.45\textwidth]{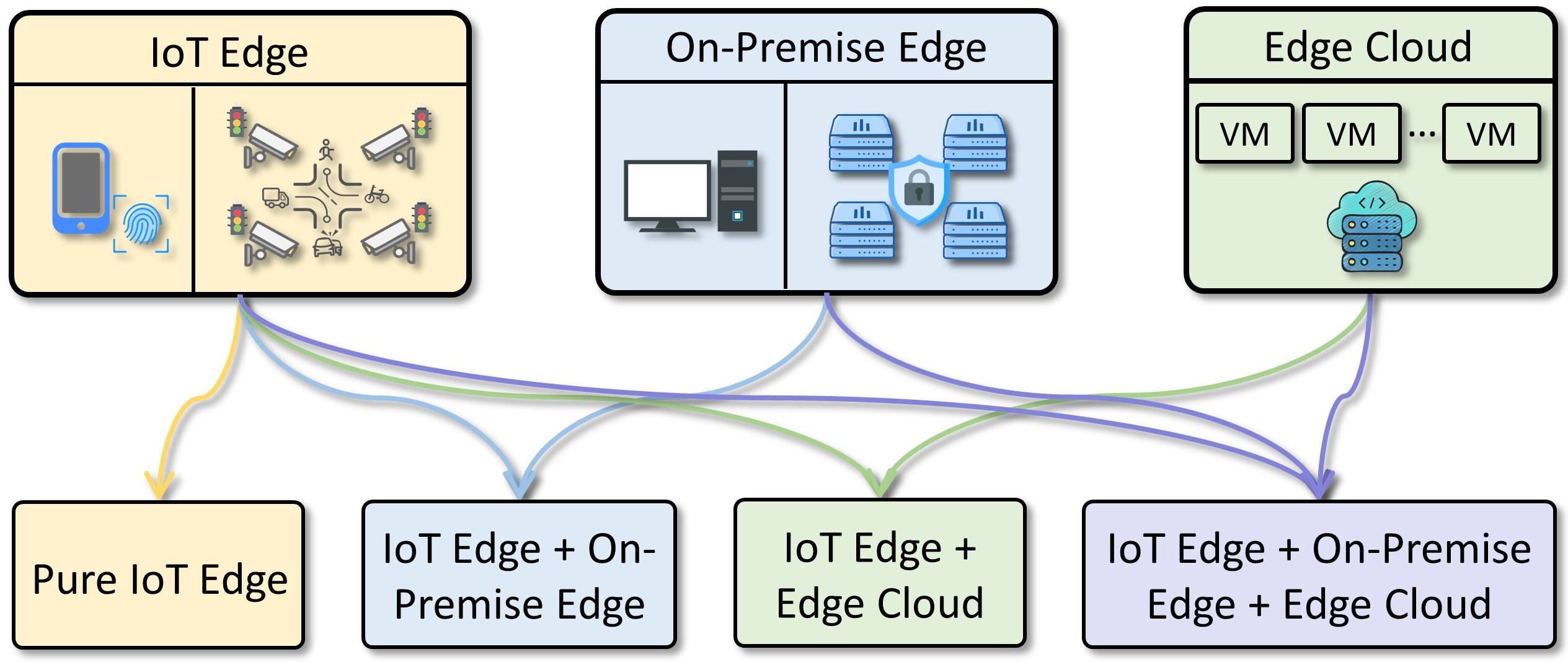}
\caption{Overview of the architectures of EVA systems.}
\label{EVA arch}
\end{figure}

\item \textit{IoT Edge + On-Premise Edge}: Since on-premise edges are close to IoT edges and are sometimes wired together, the communication overhead is minor. As a result, workloads can be offloaded from IoT edges to on-premise edges. This offloading can either be full or partial, depending on the application architecture and what trade-off needs to be hit between accuracy and latency. In addition, larger DNNs can be considered as more computational power is available at on-premise edges. The size of on-premise edges can vary from a single device (e.g., a single computer) to multiple devices (e.g., a computer cluster). Hence, in multi-device cases, workload placement is a major concern.
\item \textit{IoT Edge + Edge Cloud}: The edge cloud is another option for workload offloading. As edge clouds are only accessible via WANs and are not adjacent to IoT edges like on-premise edges, communication latency is higher. However, edge cloud outperforms on-premises edge in terms of its scalability and elasticity. Edge clouds can easily cope with dynamic loads by increasing or shrinking the amount of physical resources provisioned to VMs, which can significantly increase resource utilization \cite{VideoStorm}.
\item \textit{IoT Edge + On-Premise Edge + Edge Cloud}: This architecture is becoming prevailing and has recently been recognized by Microsoft as the only feasible approach to meeting the strict real-time requirements of large-scale LVA \cite{killerapp}. By pooling heterogeneous resources across different hierarchies and taking into account their respective strengths, the system has more opportunities to hit the best trade-off between accuracy and resource consumption. In this architecture, major cloud service providers typically offer dedicated edge devices and custom software to accomplish efficient edge-cloud collaboration. One notable example is Microsoft Rocket \cite{rocket,nearmiss,visionzero,rocketpipeline}, a configurable EVA system deployed in Bellevue, Washington, for active traffic intersection monitoring. The system employs Azure Data Box Edge to process video feeds from traffic cameras at the network edge, and uses the Azure cloud as backup computing power. Similarly, Microsoft FarmBeats \cite{farmbeats} deploys edge devices at farmers' residences to process farmland videos captured by cameras in the field. Only part of the data is forwarded to the Azure cloud for additional analysis and storage.
\end{enumerate}

\subsection{Basic Components of Edge Video Analytics Systems}
\begin{figure}[tb]
\centering
\includegraphics[width=0.43\textwidth]{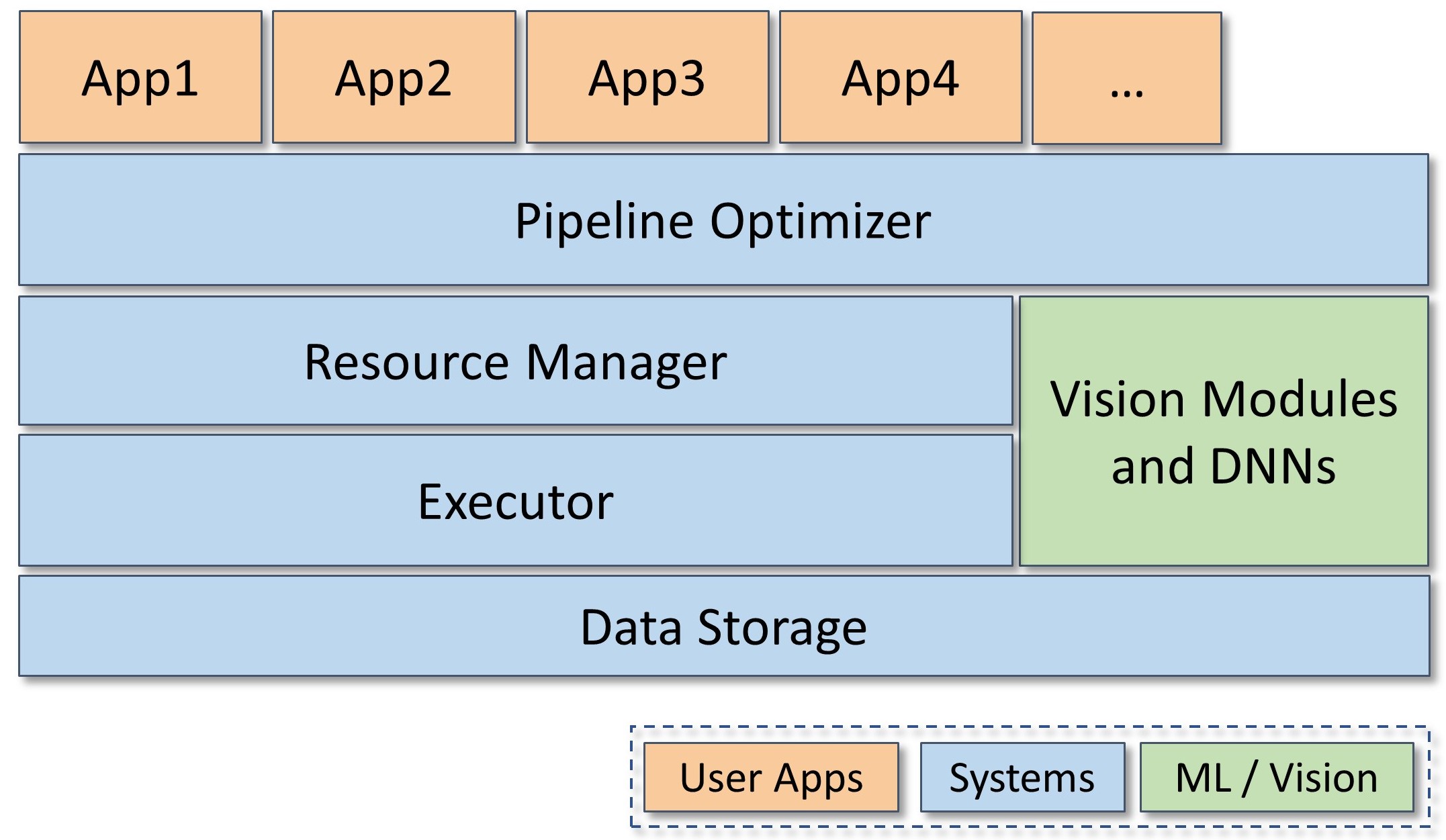}
\caption{Basic components of EVA systems \cite{killerapp}.}
\label{VAP stack}
\end{figure}

As shown in Fig. \ref{VAP stack}, a typical EVA system is composed of four essential building blocks: 1) pipeline optimizer, 2) resource manager, 3) executor, and 4) video storage. These foundational system components serve as the basis for the construction of various EVA applications, ensuring their optimal functionality and performance.

\begin{enumerate}[wide]
\item \textit{Pipeline Optimizer}: A video pipeline optimizer converts high-level video queries into video processing pipelines composed of several vision modules. Each module implements predefined interfaces to receive and process events (or data) and then sends its results downstream. Each module has its associated implementation and configurable parameters (known as \textit{knobs}) \cite{videoEdge,Chameleon,VideoStorm}. For instance, object detection can be implemented in two approaches: a) CNN-based approaches, such as YOLO; or b) traditional CV approaches, such as background subtraction. Tunable parameters include frame resolution, frame rate, model and other internal algorithmic parameters. A particular combination of implementations and knob settings is called a \textit{configuration}. Different configurations can have different performance-resource trade-offs.

To estimate the resource consumption and execution performance of each module, a profiler can be employed by the pipeline optimizer. Performance profiling is a popular technique to improve program performance based on its behaviour. It can capture program behaviour from previous runs to guide optimization decisions for future runs \cite{Profiling}. Specifically, for each vision module in the pipeline, a profiler collects the execution information (e.g., accuracy, latency, energy consumption, etc.), based on prepared representative datasets or the initial fractions of videos \cite{videoEdge, VideoStorm}. The pipeline and its generated profiles are then submitted to the resource manager, where optimization decisions are made.

\item \textit{Resource Manager}: A service level agreement (SLA) defines the level of service that users expect from a service provider, laying out the metrics to measure the service, as well as remedies or penalties to be applied if the agreed service levels are not achieved \cite{SLA,SLA2}. With the objective of minimizing resource consumption and maintaining SLAs, a resource manager determines the best configurations and placements for all pipeline modules by jointly considering the availability of resources and pipeline profiles \cite{videoEdge,VideoStorm}.

\item \textit{Executor}: The pipeline will then be executed by a suitable executor, or a group of executors if the pipeline is partitioned. The executors can be co-located (i.e. centralized in a data center) \cite{VideoStorm} or distributed (e.g., across remote devices) \cite{videoEdge}. An agent is located with the executors to continuously observe their working status. Any underutilization or overutilization of resources will be reported to the resource manager.

\item \textit{Data Storage}: Finally, raw video data will be optionally stored, and the corresponding processing results will be stored as \textit{metadata} for users’ future retrospective queries. Metadata is defined as the data that can provide information about one or more aspects of raw video data \cite{metadata1,Edge-stream}; it is used to summarize fundamental information about data, which can facilitate the tracking and manipulation of specific data. The reason why metadata is essential for VA is that machines or computers cannot “watch” videos and interpret them like humans, and metadata can make them “machine understandable” by using absolute and measurable “identifiers” such as time, location, movement, identity, size, gender, age, color, etc. These “identifiers” allow rapid retrieval of video clips containing significant changes, suspicious trends or specific events without re-processing the entire video data \cite{metadata2}. This vastly increases the efficiency of RVA, whose goal is to perform a quick retrieval of video data based on users’ queries.
\end{enumerate}

\subsection{Performance Indicators}
QoE and quality of service (QoS) are two important performance measures in an EVA system. QoE refers to the subjective quality of user experiences when interacting with a system. It can be used to measure how well the system is able to meet the needs and expectations of its users. For example, a user may expect the system to be responsive when answering queries of target objects. If the system is not able to meet these expectations, the user's QoE may be negatively impacted. QoS, on the other hand, refers to the objective performance of the system in terms of its ability to meet specific requirements or SLAs. It can be used to measure how well the system is able to deliver VA services quantitatively. For example, the system may be required to process a certain number of video streams per second with a specific level of accuracy and latency. Next, we focus on objective measures of EVA performance:

\begin{enumerate}[wide]
\item \textit{Accuracy}: We use the term “accuracy” to denote the metric that measures the analytics quality of an EVA system. Higher accuracy means that the system can produce more reliable results. This is crucial for applications like security surveillance, where unreliable analytics results could potentially trigger a false alarm. There are two types of measures. The first is to directly measure application-level metrics, which can vary across applications. For instance, in vehicle counting applications, accuracy may refer to the ratio of the estimated vehicle counts to the real vehicle counts within a given period. The second is to use task-level metrics, e.g., top-1 or top-5 accuracy for object classification tasks \cite{imagenet}, mean average precision (mAP) for object detection tasks \cite{ODSurvey}, multiple object tracking accuracy (MOTA) for MOT tasks \cite{motsurvey}, etc. Most works in this survey utilize the second one, since 1) their solutions are general and can be applied to any application containing the proposed modules and 2) the task-level metrics can more or less reflect the application performances \cite{Chameleon}.

\item \textit{Latency}: End-to-end latency refers to the actual amount of time spent on completing the entire pipeline, including execution time and data transmission time \cite{deepscale}. For real-time applications (e.g., autonomous driving and robotic surgery), the latency requirement is really stringent (e.g., less than 30 ms). Latency can be affected by many factors, i.e., the internal factor: the arithmetic power of executors (i.e., CPUs, GPUs, FPGAs, and ASICs) and the external factor: the resource availability (i.e. computation and network) of executors. We consider resource availability an external factor because it can change based on the emergence of external workloads. Variants based on latency are also introduced for different evaluation purposes, e.g., deadline missing rate \cite{RT-mDL} and latency service level objective (SLO) missing rate \cite{Distream}. 

\item \textit{Throughput}: Originally, throughput was a measure of how many units of information a system can process in a given amount of time. In the VA domain, it is introduced to quantify how many frames an EVA system can process in a certain amount of time, and determine if the system can achieve real-time processing. Live EVA systems need to process streaming video frames in a continuous manner, and having a high throughput is essential to keeping up with the incoming video streams. For a single-executor EVA system, throughput is almost equivalent to latency, while for a multi-executor EVA system, throughput can quantify the computing power of the entire system \cite{Distream}. Typically, throughput is defined as “the number of requests (or frames) processed per second” \cite{kalmia}, but it can vary based on specific systems. For instance, in Distream \cite{Distream}, throughput is measured as “the number of inferences processed per second (IPS)”, while in CEVAS \cite{CEVAS} throughput represents the ratio of the analysis frame rate to the frame rate of the video stream (e.g., for a video stream with a frame rate of 10 FPS, a throughput of 50\% indicates that the video stream is analyzed at 5 FPS).
\end{enumerate}

\begin{figure}[tb]
\centering
\includegraphics[width=0.43\textwidth]{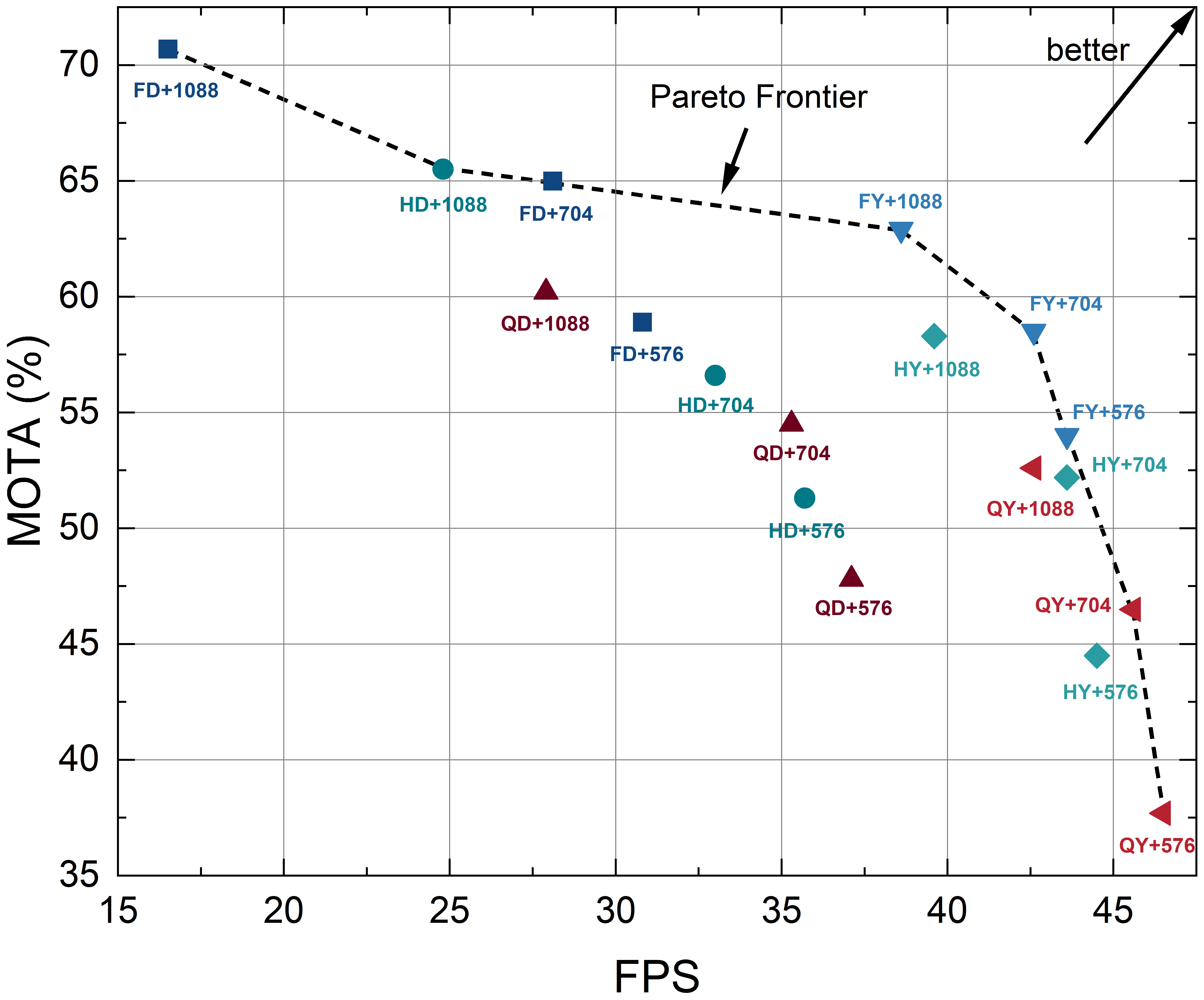}
\caption{Pareto frontier (black dashed curve) of the pipeline. 1088, 704 and 576 are short for $1088\times608$, $704\times384$, $576\times320$, respectively. FD, HD, QD, FY, HY and QY are short for Full-, Half-, Quarter-DLA-34 and YOLO, respectively.}
\label{pareto}
\end{figure}

Due to the non-trivial trade-off between performance and latency in EVA, one cannot optimize all metrics simultaneously. A Pareto frontier characterizes a set of optimal solutions in a multi-objective optimization problem, where no further improvement in one metric can be achieved without compromising at least one of the other metrics. For example, consider FairMOT \cite{fairmot}, a one-shot MOT tracker. The pipeline has several knobs (i.e., frame resolutions and backbone models) to choose from. Three resolutions: \{$1088\times608$, $704\times384$, $576\times320$\} px, and six models: Full-, Half-, Quarter-DLA-34, Full-, Half-, Quarter-YOLO are considered. Resource-intensive configurations (e.g., $1088\times608$ px+Full-DLA-34) typically lead to better accuracy at the cost of higher latency, and vice versa. The accuracy-speed trade-off of the pipeline is illustrated in Fig. \ref{pareto}, where the data points represent the results achieved with different configurations. The black dashed curve is the Pareto frontier. The points on the Pareto frontier represent a set of optimal configurations with respect to different application requirements. If the application requires a MOTA greater than 65\%, $704\times384$ px+Full-DLA-34 is the best choice, since it produces the highest FPS while satisfying the accuracy requirement. In contrast, if the application demands an FPS of over 45, $704\times384$ px+Quarter-DLA-34 is the best option as it meets the FPS requirement while maintaining the highest possible accuracy.

\subsection{Application Architectures}
\begin{figure*}[tb]
\centering
\subfloat[]{
\includegraphics[width=0.3\textwidth]{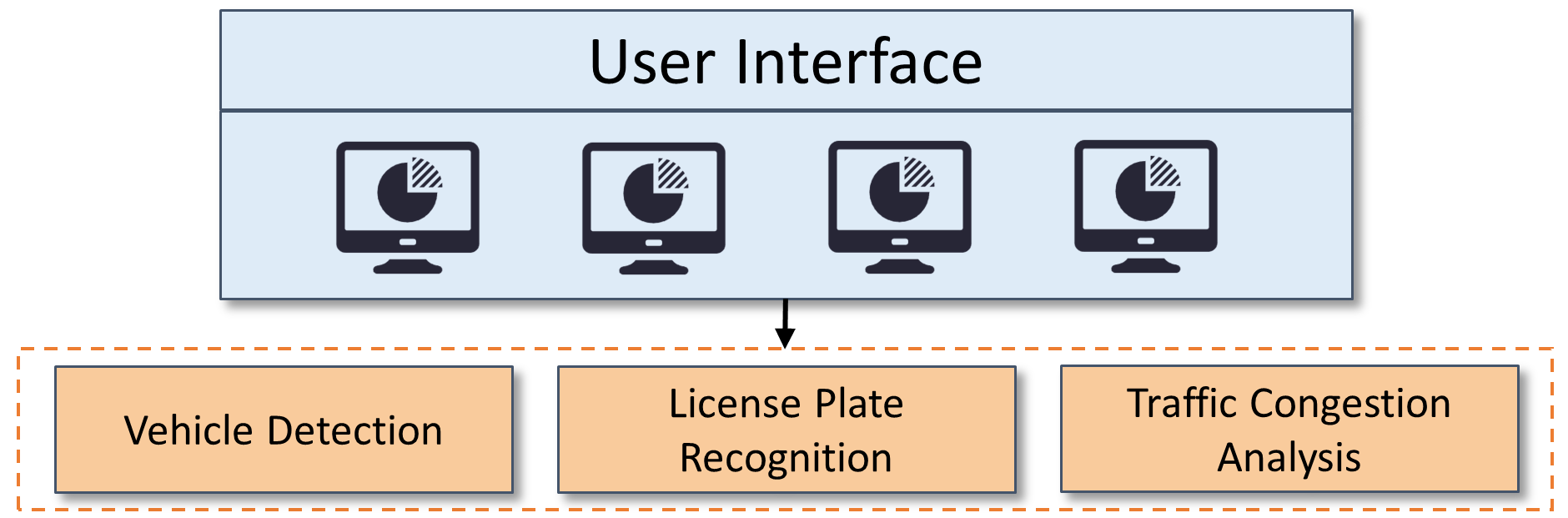}
\label{monolithic}}
\hfil
\subfloat[]{
\includegraphics[width=0.3\textwidth]{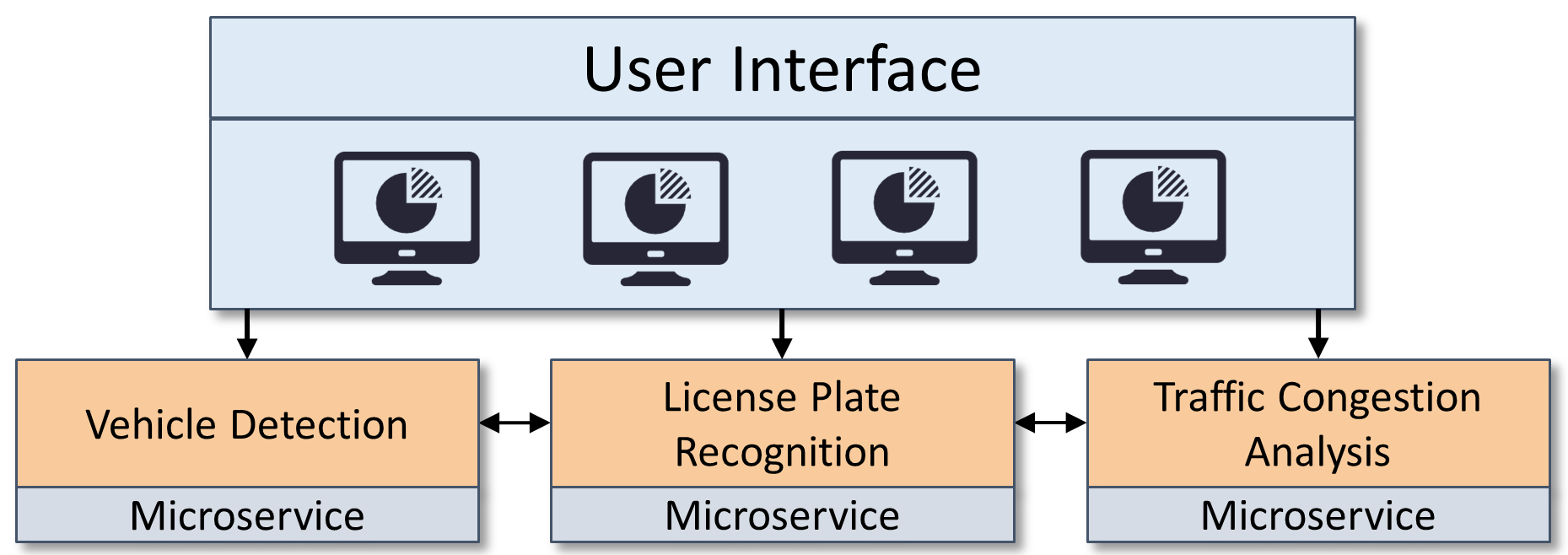}
\label{microservices}}
\hfil
\subfloat[]{
\includegraphics[width=0.3\textwidth]{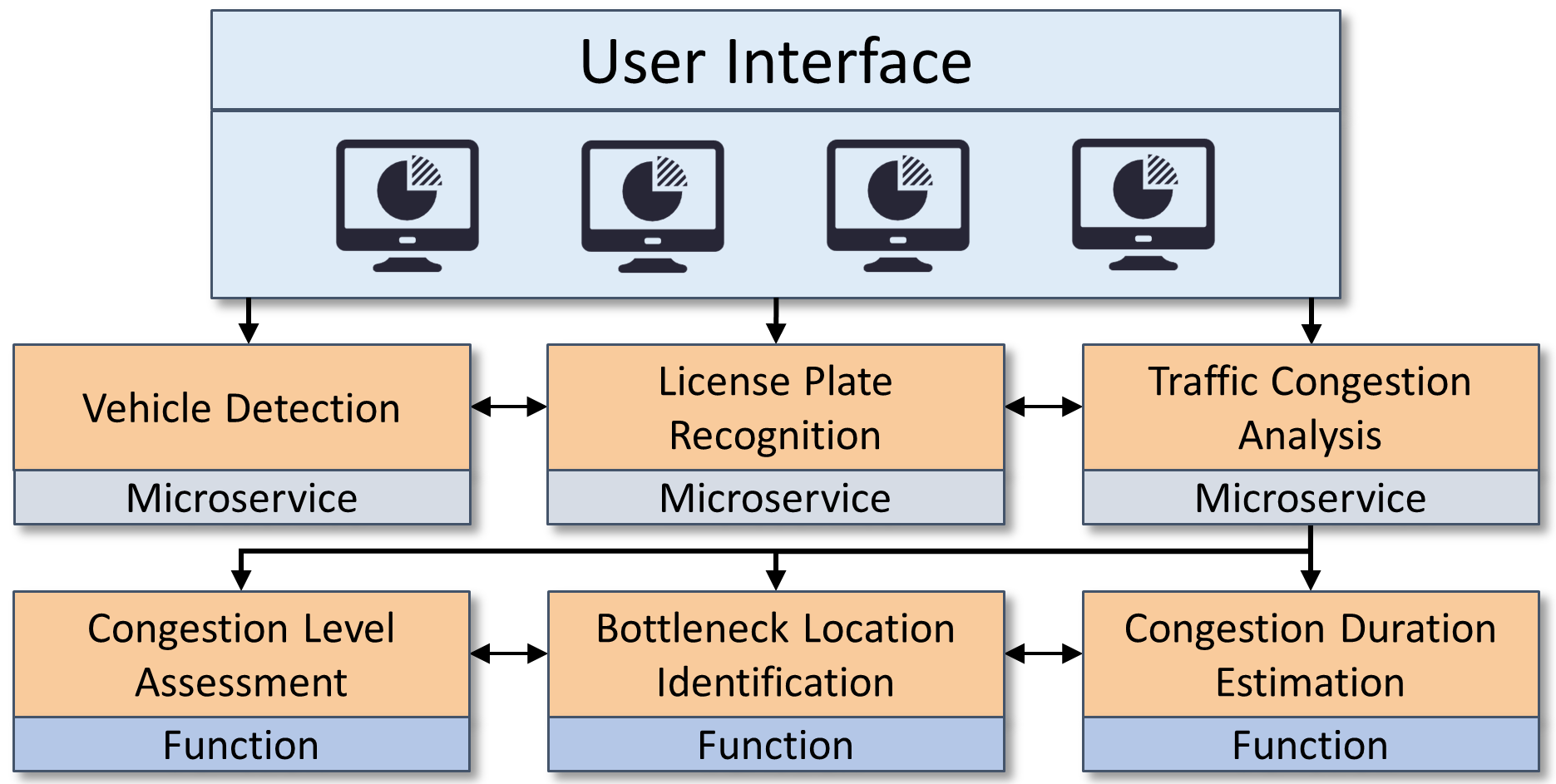}
\label{serverless microservices}}
\caption{Three types of application architectures. (a) Monolithic architecture. (b) Microservices architecture. (c) Serverless microservices architecture.}
\label{app arch}
\end{figure*}

As shown in Fig. \ref{VAP stack}, applications are implemented based on the system components. The design of an application is highly dependent on the application requirements, e.g. scalability, flexibility, etc., and specific systems, e.g., the number of executors, the resource availability, etc. To optimize the application performance and alleviate future maintenance difficulty, the application architectures have experienced several evolutions: from monolithic architectures to microservices architectures, and finally to serverless microservices architectures.
\begin{enumerate}[wide]
    \item \textit{Monolithic Architecture}: Traditionally, all application processes are tightly coupled and run as a single service, i.e., in the \textit{monolithic architecture} \cite{mm1,mm2}. This means that if one application module experiences a spike in demand, the entire architecture needs to be scaled, i.e., by provisioning a VM to host a new application instance. Adding or improving the features of a monolithic application becomes more complex as its code base grows. The monolithic architecture also adds risk to service availability because the dependency and coupling among modules amplify the impact of a single module failure. For instance, consider an EVA application for monitoring a city's traffic conditions, as shown in Fig. \ref{monolithic}. If this application uses a monolithic architecture and experiences an unexpected increase in video feeds from various intersections, the vehicle detection, license plate recognition, and traffic congestion analysis services may all experience slow responses, errors or even downtime. This is because the entire application is contained within a single codebase and must be deployed and scaled as a single unit.
    
    \item \textit{Microservices Architecture}: With the microservices architecture, an application is built with a collection of loosely-coupled fine-grained \textit{microservices} \cite{microservices,mm1,mm2,msexample}. These microservices can be independently developed in different programming languages, separately deployed in different infrastructures, and managed by different operation teams. They communicate with each other via well-defined and lightweight application programming interfaces (APIs). Returning to the previous example, by adopting a microservices architecture, the application could be decoupled into three separate microservices, as shown in Fig. \ref{microservices}. If the vehicle detection service experiences a surge in video feeds, it can be scaled independently, while the other microservices continue to operate normally.

    The microservices architecture has been widely adopted in real-world EVA applications due to its flexibility and scalability. In Microsoft Rocket \cite{rocket,nearmiss,visionzero,rocketpipeline}, the application pipeline is composed of a series of separate microservices: decoding, background subtraction, light DNN object detector, heavy DNN object detector and database. Each of them can be independently configured to fit different application requirements and to execute over a distributed infrastructure, potentially spanning specialized edge hardware (e.g., Azure Data Box Edge) and Azure cloud (e.g., Azure Machine Learning and Cognitive Services). For instance, if the edge device does not have sufficient resources to hold two object detectors, the heavy one could be placed in the cloud and only be invoked when the light one cannot produce reliable results. Similarly, the NVIDIA multi-camera smart garage application \cite{deepstream} splits the entire pipeline between the edge and the cloud. At the edge side, a self-contained microservice performs video decoding, detection, global positioning and single-camera tracking on the video feeds from single cameras. The metadata is then sent to the cloud, where a multi-camera tracker will aggregate the data, and an analytics engine will detect events, including anomalies, garage occupancy and traffic flow. When dealing with a growing number of cameras, only the microservice at the edge side needs to be scaled up.
    
    \item \textit{Serverless Microservices Architecture}: Serverless computing is a cloud computing model where a cloud provider manages the infrastructure required to run and scale applications. Tasks such as load balancing, capacity planning, auto-scaling, fault tolerance, storage, networking, security, maintenance, etc., are handled by the cloud provider, whereas application developers are only concerned with developing code.
    
    In a serverless architecture, applications are broken down into smaller, independent functions (e.g., a few lines of code) that can be executed on demand, rather than continuously running on a server \cite{applicationserverless2}. These functions are triggered by specific events, such as an HTTP request or a change in a database, and the cloud provider allocates the necessary resources to execute the functions \cite{applicationserverless2}. The resources are released after the execution is complete.
    
    Microservices architecture can be combined with serverless computing \cite{applicationserverless,applicationserverless2}. The resulting architecture, \textit{serverless microservices architecture} \cite{serverlessmicroservice} is widely adopted in commercial solutions. In a serverless microservices architecture, a microservice can be implemented as a set of event-driven functions. As shown in Fig. \ref{serverless microservices}, the traffic congestion analysis microservice could be divided into serverless functions for assessing congestion levels, identifying bottleneck locations, and estimating the duration of congestion. These functions are triggered by specific events, such as the output of vehicle detection or license plate recognition. Overall, this architecture provides better granularity and resource utilization. However, the choice between a microservices architecture and a serverless microservices architecture depends on the specific needs and requirements of an EVA application.
\end{enumerate}

\section{Enabling Techniques in Edge Video Analytics Systems}
In this section, we introduce key enabling techniques that optimize the real-time performance of EVA systems.

\subsection{Performance Profiling}
The performance of an application is determined by many factors, such as configuration and workload placement. As explained in Section V-C, different configurations can bring different resource-accuracy trade-offs. Choosing a good configuration can minimize resource demands while maintaining analytics quality. Workload placement is another important problem in edge computing since application performance can vary a lot based on the resource availability of executors. Performance profiling gathers information regarding the program characteristics during execution, and is a popular technique to reason about the dynamic behaviour of a program \cite{Profiling}. The profiling information (known as \textit{profile}) can then be used to optimize configuration and placement decisions. In general, profiling can be done either offline or online.
\begin{enumerate}[wide]
    \item \textit{Offline Profiling}: To get an accurate profile, a natural approach is to do a one-time but exhaustive offline profiling. A profiler executes instrumented tasks multiple times with different inputs and configurations on all possible executors and outputs metrics such as accuracy, execution time, resource demand, memory footprint, input or output data size, etc. The datasets used in profiling tasks need to be representative of target application scenarios. Typically, the initial fraction (usually several minutes) of videos is labelled by detectors with “golden” configurations \cite{Chameleon,videoEdge,VideoStorm}, which can be computationally expensive but are known to produce high-quality results, and then utilized for profiling. The profiling cost can be prohibitive since the search space tends to grow exponentially with the number of parameters (e.g., configurations, placements, etc.). For instance, VideoEdge \cite{videoEdge} considers a search space of 1800 combinations from five resolutions, five frame rates, three object detector implementations, four tracker types and six placements for trackers. Similarly, VideoStorm \cite{VideoStorm} considers a search space of 414 combinations, and it takes 20 CPU days to generate the profile for a 10-minute video. The situation is even worse for lengthier videos. Even though parallelism \cite{VideoStorm,AWStream} has been exploited to accelerate profiling, the high resource demand for exhaustive profiling remains a challenge. To alleviate this problem, VideoEdge \cite{videoEdge} merges common components among multiple configurations and caches intermediate results. For instance, assume that both components in the pipeline $A\rightarrow B$ have two implementations: $A_1, A_2$ and $B_1, B_2$. Four implementation plans have to be profiled: $B_1A_1, B_1A_2, B_2A_1$ and $B_2A_2$. If profiling is done natively, $B_1$ and $B_2$ will run twice on the same video data. So, merging common components $B_1$ and $B_2$ and caching their results can avoid redundant runs. Another way to reduce profiling costs is to sub-sample the configuration space. ApproxDet \cite{ApproxDet} only profiles 20\% of the configurations, at the cost of generating a less accurate profile \cite{AWStream}.
    
    \indent The inability to capture varying visual characteristics in real scenes is another problem of offline profiling. As a result, the decision may fail to account for the intrinsic resource-accuracy trade-offs, leading to either resource waste from unnecessarily expensive configurations or SLA violations. Take vehicle detection as an example, a configuration with low resolution (e.g., 480p) and frame rate (e.g., 5 FPS) is sufficient to retain acceptable accuracy if cars are moving slowly, e.g., at a traffic stop. This configuration may fail if cars move fast. A promising solution to this problem is content-aware profiling, which incorporates content features as additional dimensions of the profile. Then, a prediction model is trained based on the profile and used to estimate the performance online. In this way, the model can adapt at runtime since the video characteristics are involved in the prediction. Various content features are considered in different approaches. ApproxNet \cite{ApproxNet} uses edge values to measure the frame complexity and ApproxDet \cite{ApproxDet} uses the number of objects, their sizes and moving speeds to characterize their moving patterns. Other heavy-weight features such as Histogram of Colors (HoC), HOG and CNN-based features (e.g., feature vectors extracted by lightweight CNNs) are used in LightReconfig \cite{LiteReconfig} and SmartAdapt \cite{SmartAdapt} in offline profiling. CEVAS \cite{CEVAS} quantifies video content dynamics by self-defined metrics: average number of objects per frame and average number of unique objects per frame.

    \item \textit{Online Profiling}: One weakness of the aforementioned approaches in incorporating scene dynamics in offline profiling is that the selection of features highly relies on domain expertise. Moreover, it is hard to ensure the generalizability of such handcrafted and low-level features \cite{infi}. They may perform well in specific scenarios but not in others. Designing good and universal content features is a challenging task. Recently, online profiling has become a popular alternative. Different from offline profiling, which is done once or infrequently (e.g., once a day \cite{Chameleon}), online profiling updates the profile periodically (e.g., every few seconds or minutes \cite{Chameleon}) during video streaming. The main challenge of online profiling is how to reduce the overhead of periodic profiling. As mentioned earlier, naively performing a one-time profiling can take a large amount of time, let alone periodic ones. To mitigate this challenge, Chameleon \cite{Chameleon} first does full online profiling to exhaustively profile all configurations and generate several candidate configurations. Then, by leveraging cross-camera correlation and video content consistency, the candidate configurations are shared and propagated both spatially and temporally. Thus, the significant cost of full profiling can be amortized. Knob independence (i.e., for a given knob, the relationship between its resource and accuracy is independent of the values of the other knobs) is also leveraged to reduce the search space from exponential to linear. Notably, this solution is coarse-grained and relies on setting a pre-fixed time interval for profiling. Alternatively, techniques like scene understanding can be used to detect scene changes, predict the need for re-profiling and trigger fresh profiling, thus saving the profiling cost even further. However, existing studies regarding this aspect are scant. AWStream \cite{AWStream} considers a combination of offline profiling and online profiling. Online profiling is used to gradually refine the bootstrap profile obtained offline. To speed up, AWStream only profiles a subset of configurations, which are Pareto-optimal (i.e., the ones on the Pareto boundary). A full profiling is triggered to update the current profile only when more resources are available.
\end{enumerate}

\subsection{Input Filtering}
Input filtering aims to remove redundant computation, and thus save resource consumption. The type and extent of redundant computation can vary across application scenarios; it can refer to stationary or irrelevant frames \cite{Glimpse,FastQ,FilterForward,infi}, redundant inferences \cite{deepcache,infi}, uninformative or duplicated regions \cite{CrossRoI,REMIX,DeepStream2}, etc. Input filtering is realized by several techniques discussed in detail next.
\begin{enumerate}[wide]
    \item \textit{Pixel-Level Difference Detector}: Glimpse \cite{Glimpse} applies a pixel-level frame difference detector to detect scene changes. Only when the number of “significantly different pixels” between two consecutive frames exceeds a threshold will the current frame, called the trigger frame, be sent to an edge server for assistance. ApproxNet \cite{ApproxNet} also utilizes a scene change detector to initiate the succeeding frame complexity estimator, but unlike Glimpse, it utilizes changes in the color histogram (R channel) of pixels. Similarly, ClougSeg \cite{CloudSeg} proposes a two-level thresholding method based on pixel deviation, where the lower threshold is for selecting useful frames and the higher one is for key frames (i.e., the most useful frames). The heavy inference will only be imposed on key frames. Notably, instead of using pre-defined thresholds that may fail due to dynamic video content, CloudSeg adapts the thresholds according to network conditions and application requirements. Reducto \cite{Reducto} is an extension of CloudSeg. It dynamically adapts filtering configurations (i.e., the feature type of the difference detector and the filtering threshold) for different queries and videos by efficiently coordinating with the server. Specifically, Reducto selects the best feature type using offline server profiling, and predicts the filtering threshold with a lightweight regression model. Periodical retraining is also considered in case the model is outdated.
    \item \textit{Binary Classifier}: Binary classifiers can be utilized in input filtering to determine redundant frames and frames to be retained. FilterForward \cite{FilterForward} identifies the most relevant video clips to the applications using microclassifiers. Wang et al. \cite{BLVA} propose a cascaded filter composed of two classifiers: EarlyDiscard and Just-in-time-Learning (JITL). EarlyDiscard is a customized CNN classifier for selecting mission-specific video frames, while JITL, which is periodically trained based on the frames reported by the EarlyDiscard filter, is used to verify and further eliminate wrong results. FFS-VA \cite{FFS-VA} presents a similar multi-stage filtering method, which exploits a difference detector, a CNN classifier, and an object detector to progressively eliminate background frames, non-relevant frames and frames with few target objects.  
    \item \textit{Inference Result Reusing}: There is much redundancy when performing inferences on continuous video streams, as the inference results of previous frames can be reused. DeepCache \cite{deepcache} finds that only a small portion of the content changes in consecutive frames. Hence, it caches the inference results of previous frames, and identifies redundant blocks using block-wise matching. The inference results of these blocks are then reused over time. Similarly, BlockCopy \cite{BlockCopy} employs deep reinforcement learning (DRL) to determine the informative blocks of the current frame, based on information from the last frame. Once the informative blocks are decided, block-sparse convolution, originally proposed in \cite{SegBlocks}, is used to process the current frame. Specifically, the identified informative blocks are extracted and batched for inference, while the features of uninformative ones are copied from previous executions. One limitation of BlockCopy is that it cannot reuse previous computations for objects that move across the scene, even when corresponding visual features do not change much. InFi \cite{infi} proposes an end-to-end learnable input filter, which can predict the redundancy score of the input and perform filtering in two manners: SKIP and REUSE. SKIP aims to filter irrelevant frames (i.e., frames without target objects), and REUSE attempts to filter frames whose results can reuse the previously cached inference results. Due to end-to-end learnability, InFi can perform robust filtering in a workload-agnostic manner. Notably, InFi also provides a generic formalization of the input filtering problem and a theoretical filterability analysis. FDDIA \cite{FDDIA} reuses the detection results of the previous frame to determine the RoIs in the current frame. The RoIs are then down-sampled and integrated into a new image for batch processing. Similarly, AccDecoder \cite{AccDecoder} uses DRL to adaptively select key frames for quality enhancement and inference. It then reuses the information from these enhanced frames, such as the cached inference results, motion vectors, and residuals, to eliminate redundant computations on the remaining frames. 
    \item \textit{Spatial Filtering Mask}: Different from the above approaches, which eliminate temporal redundancy, spatial filtering aims to remove spatial redundancy, i.e., duplicated regions across cameras. A representative work is CrossRoI \cite{CrossRoI}, which exploits the intrinsic physical correlations of cross-camera viewing fields to avoid duplicated computation for the same objects. Specifically, CrossRoI operates in two distinct phases: an online phase and an offline phase. In the offline phase, it establishes the cross-camera correlation using object re-ID algorithms and generates the corresponding RoI mask for each camera. In the online phase, cameras only keep data in these regions covered by the RoI masks and run a specialized RoI-based object detector to reduce inference time. One drawback of CrossROI is that each camera only focuses on the assigned regions while being blind to the others, losing the chance to improve the detection results by leveraging other camera views. Polly \cite{Polly} tackles this issue by introducing a confidence-aware method. The regions with high sharing confidence are shared and mapped across cameras, while the ones with low confidence are patched for further inference. It is worth noting that spatial filtering is orthogonal to temporal filtering, so the two can be combined together to achieve further computation savings.
\end{enumerate}

\subsection{Hierarchical Inference}
It is known that DNN model inference is compute-intensive. Deeper network architectures bring higher accuracy but also incur more computation overhead. Naively running a deep model on a video can waste resources, as not every frame is difficult and requires such a heavy model. A hierarchical architecture allows more flexibility in model inference, i.e., using a lightweight model to handle “easy” frames and a heavy model to handle challenging frames. Two techniques can be employed to enable hierarchical inference:
\begin{enumerate}[wide]
    \item \textit{Model Early Exit}: Early exit first appeared in BranchyNet \cite{BranchyNet}, a Dynamic DNN (DDNN) model proposed in 2016. In contrast to traditional DNNs, DDNNs \cite{DDNN,ibranchy} are capable of performing conditional computations and selectively activating a subset of the network model, whereas traditional DNNs use the entire network in the computation even when a certain portion of the network is sufficient to make a good inference. BranchyNet supports the early inference of certain input samples using a multi-branch and multi-exit design. Like a traditional DNN classifier, the BranchyNet network architecture consists of a multi-layer network followed by a softmax layer for output predictions. However, in addition to the main network, a small network called branches is added to the outputs of different layers of the main network. These branches, similar to the main network are also followed by a softmax layer. These outputs, as well as those of the main network, are called \textit{exits}. The multi-exit approach implemented in BranchyNet is based on the observation that the earlier layers of the network can perform inference for most input samples accurately. With early exits, the average runtime can be reduced. Specifically, during the inference phase, BranchyNet computes the entropy of the softmax output at an exit. If the entropy of the input sample is larger than the given threshold, the sample is sent to the next exit for inference, and the process continues till it reaches the final exit at the end of the main network; otherwise, the softmax output is taken as the model prediction.

    Recently, several works have utilized the early exit technique and DDNNs to achieve low-latency and efficient VA on edge devices. ApproxNet \cite{ApproxNet} designs a DDNN classifier based on BranchyNet to perform multi-class object classification. The network architecture is composed of six \textit{stacks} and each stack has four or six ResNet layers and a variable number of blocks from the original ResNet design. These stacks are connected to six exits, each of which consists of a spatial pyramid pooling (SPP) module, a fully connected (FC) layer, and a softmax layer. Via these exits, inference can terminate earlier without executing the entire network. Exits can be selected based on the resource availability on the device, the content characteristics and application requirements to achieve a desired trade-off between accuracy and latency. In addition, switching exits is much faster than switching DNN models, as loading a new DNN model into memory is time-consuming (e.g., up to 1 second \cite{Chameleon}). EdgeML \cite{EdgeML} combines DDNN with partial offloading. In the offline stage, a DDNN model is constructed by inserting branches across the layers of the original DNN model, and then trained and fine-tuned. In the online stage, a DRL-based optimizer is utilized to determine the optimal execution policy, including the layer-wise partition point and the threshold for the early exit of each branch, based on the available network bandwidth, input data characteristics, and the user's latency and energy requirements. Then, the model is partitioned and separately executed by the edge and cloud. A similar problem is considered in MAMO \cite{MAMO}, but it jointly considers the resource allocation of containers. One limitation of the aforementioned works is that they are input-agnostic, as the number and locations of exits are pre-determined based on heuristics or domain expertise. A bad setting can potentially diminish the benefit of early exit \cite{FlexDnn}. To address this issue, FlexDNN \cite{FlexDnn} proposes an input-aware architecture search scheme to find the optimal early exit insertion plan (i.e., the number and locations of exits) that balances the trade-off between early exit rate and its computational overhead, based on the input data. Targeting video activity recognition, FrameExit \cite{FrameExit} proposes a variant of early exit, known as frame early exit. To achieve accurate activity recognition, it is necessary to analyze multiple frames within a video segment. Gating modules in FrameExit are trained to automatically identify the earliest exit point based on the inferred complexity of the input video. This strategy allows for the processing of only a subset of frames, effectively reducing computational overhead without compromising recognition accuracy.
    
    \item \textit{Model Cascade}: The key idea of the model cascade is similar to the model early exit, i.e., running models of different complexity on different frames, but the models are separated rather than integrated as in a DDNN. One line of work uses a small model to handle input frames and only when the results are not reliable (e.g., if the results have a low confidence score \cite{flexpatch,rec}, or anomalies are detected \cite{marvel,edgesharing}), a heavier model will be invoked for further processing \cite{edgesharing, CBO}. Therefore, the small model acts like a soft “filter” and allows for efficient processing of live video streams by avoiding resource-intensive processing on easy frames (typically with less related information). Note that this is different from input filtering, where non-informative frames are eliminated. For example, Rocket \cite{rocket} cascades a light and a heavy DNN object detector in the pipeline. Frames are processed at an edge server by the light detector unless the detection results are unreliable (i.e., the average confidence is lower than a certain threshold), in which case they will then be sent to the cloud, where the heavy detector will be called for a second-round inference. SurveilEdge \cite{SurveilEdge} utilizes a similar cascade architecture in an event-driven EVA system, where two classifiers with different complexity are placed on the edge server and the cloud. Once a query is issued, a lightweight, context- and query-specific (CQ-specific) CNN classifier is trained based on the context-specific training set and then deployed on the edge server to process the video stream and answer the query. At the same time, a highly-accurate CNN classifier is deployed in the cloud as a backup to handle the frames deemed challenging (based on pre-defined thresholds) by the edge classifier. REACT \cite{REACT} and AttTrack \cite{AttTrack} represent another line of model cascade. A lightweight model processes most of the frames at the edge, while a large model deployed in the cloud is periodically invoked for accurate detections. Information from the large model is also transferred to the edge model to enhance its detection performance. AttTrack differs from REACT in two-folds. Firstly, instead of transferring low-level information like bounding boxes, AttTrack transfers high-level knowledge (“attention”), represented in a heatmap, from the large model to the tiny model. This design allows AttTrack to use attention transfer to minimize the performance gap between the two models during training. Second, REACT operates in an asynchronous manner, allowing for potential delays in receiving cloud-based detection results, while AttTrack blocks edge operations until the heatmap from the previous key frame is received from the cloud.
\end{enumerate}

\subsection{Configuration Optimization}
As mentioned previously, a typical VAP involves several video processing components. Core components such as object tracking may have many implementation choices, but no single one is always the most accurate or efficient across all scenarios \cite{videoEdge}. Each implementation can also have several tuning knobs like resolution and frame rate. Thus, a VAP can have thousands of configurations (i.e., combinations of implementations and their knob values). The choice of configuration can impact the resource consumption and accuracy of an application \cite{videochef,videoEdge,VideoStorm,Chameleon}. For instance, using a high frame resolution or a complex DNN model in object detection enables accurate detection but also demands more computing resources. The “best” configuration can be the one with the lowest resource demand whose accuracy meets the application requirement or the one that maximizes accuracy subject to latency and resource constraints. A configuration is Pareto-optimal if one cannot unilaterally improve one metric (e.g., accuracy) without degrading the other (e.g., latency). It is non-trivial to decide the best configuration for a VAP, since it varies over time, and characterizing the trade-off between resource usage and performance is a challenging task in itself.
     
A large number of works consider adapting configurations at runtime. Knobs like video quality (e.g., resolution, frame rate, bitrate) \cite{Chameleon,videoEdge,VideoStorm,AWStream,AdaVP,deepscale,BNTL,Cuttlefish,DeepDecision,JCAB, FACT,ECQC,lmdn,leaf,edgeadaptor,EdgeVision,Maxim} and DNN model \cite{Chameleon,videoEdge,VideoStorm,ApproxDet,JCAB,SCYLLA,SmartEye,NestDnn,palleon,edgeadaptor,EdgeVision} are widely explored. In addition, camera configurations \cite{CamTuner}, e.g., brightness, contrast, color-saturation, sharpness, etc. are shown to impact the analytics quality. Furthermore, internal algorithmic parameters like model input size, filter feature type, sampling interval, number of proposals, and down-sampling ratio can impact accuracy-efficiency trade-offs as well \cite{ApproxDet,LiteReconfig,SmartAdapt}. 
    
In general, configuration optimization can be done \textit{proactively} or \textit{reactively}. 
\begin{enumerate}[wide]
    \item \textit{Proactive}: Proactive approaches make configuration adaptation decisions based on real-time system status. Hence, they require precise information about the available network bandwidth (typically done with bandwidth probing), CPU or GPU resource availability, and the expected performance under the current condition (e.g., based on a complex performance model), all in real-time. Conventional proactive approaches have three steps: 1) probing available resources (e.g., compute and network resources), 2) estimating the performance of each configuration (e.g., accuracy and latency), and 3) solving an optimization problem to maximize a certain objective function (e.g., desired accuracy, latency) given the constraints (e.g., resource, energy budget) and obtain the optimal configuration. Specifically, compute resources can be quantified by the number of CPU cores \cite{videoEdge,VideoStorm},  CPU or GPU utilization rate \cite{RT-mDL,Gemini,Hetero-edge}, contention level \cite{ApproxNet,ApproxDet}, etc.; network resources can be characterized by available bandwidth \cite{videoEdge,VideoStorm}. Profiling techniques, as introduced previously, are utilized to obtain the execution performance of different configurations. However, regardless of offline or online profiling, the entire pipeline must be executed to obtain the final result, which is inefficient. To address this issue, DeepScale \cite{deepscale} proposes a surrogate-driven approach. By using self-supervised learning, DeepScale can predict the heatmaps of different configurations (i.e., resolution) in an inference. The detection rate, which can be derived based on the heatmaps, is used as a surrogate to predict the performance of different configurations since the detection rate is highly correlated with the detection accuracy. By employing a cheap surrogate, expensive computations can be avoided since measuring the performance metric in real-time is difficult. Other proactive configuration adaptation approaches differ in the formation of objective functions and constraints. For instance, VideoEdge \cite{videoEdge} considers the joint decision of optimal configuration and workload placement by tackling a combinatorial problem. To efficiently solve the NP-hard problems, a heuristic-based approach is employed to obtain an approximate (near-optimal) solution. Similarly, JCAB \cite{JCAB} jointly optimizes the configuration and bandwidth allocation, and the original combinatorial problem is transformed into a series of one-slot optimization problems, each of which is solved by leveraging the Markov approximation and the KKT condition. The solution is proven to achieve near-optimal performance.
    
    Recently, deep reinforcement learning, a powerful sequential decision tool, has been applied in proactive configuration adaptation \cite{DRLsurvey,Cuttlefish,CamTuner,EdgeML,Elf,ADQN,MEVAO,casva,BAS,EdgeVision,Maxim}. To get the maximum accumulated rewards, a DRL agent explores and exploits the environment through randomization of actions. The key challenge in DRL-based approach is how to define states, actions and rewards, and learn good policies efficiently as the action space tends to be very large. For example, Cuttlefish associates the state space with the key factors that will affect the configuration choice, including bandwidth, moving velocity of objects, and historical configurations. The agent, trained with the Asynchronous Advantage Actor-Critic (A3C) algorithm, takes these states as input and selects the optimal action (configuration) in the action space. The reward function is carefully crafted by jointly considering three metrics that can directly reflect users’ QoE, i.e., detection latency, accuracy and fluency of video play. Similarly, Gemini \cite{Gemini} takes the estimated bandwidth, resource utilization (i.e., CPU and GPU), delay requirement, and frame rate as input, and formulates a multi-armed bandit problem to solve for the optimal CPU-GPU time partition of the FPGA, and choice of video resolution and DNN model. The reward is defined as the uncertainty of prediction results since accuracy cannot be measured without ground truth.
    
    \item \textit{Reactive}: In contrast to proactive approaches, which can be viewed as open-loop solutions to configuration selection, reactive approaches adapt configurations based on the differences from performance targets in a close-looped manner. Nigade et al. \cite{BNTL} present a prototype model that guarantees strict SLOs for VAPs through feedback control-based resolution adaptation. The absolute difference between the real latency from the last frame and the target latency serves as feedback. A threshold-based controller takes the feedback and decides the resolution for upcoming frames. Although the threshold is empirically set and thus unlikely to work well in all scenarios, the work is among the first to demonstrate the value of reactive approaches. Reactive approaches do not require precise information about the available resources, which are highly dynamic and hard to obtain on time. There is no need to probe resource availability and explicitly model the relationship between performance and resources. But the simplicity comes at the cost of longer or lack of convergence and suboptimality. 
\end{enumerate}

\subsection{Workload Placement and Scheduling}
In an EVA system, users can submit their requests remotely. A request can take many forms based on the specific application: it could be a query issued by a user (i.e., event-driven VA) or a video feed continuously streamed from a camera (i.e., continuous VA). The service provider is responsible for fulfilling these requests and guaranteeing SLAs. To manage network, computation and storage resources efficiently, the service provider needs to determine the optimal devices to handle these requests. As mentioned in Section V-E, an application can be implemented in the microservices architecture, where each component of the pipeline can be implemented as a separate microservice to be deployed, scaled and updated independently. We call a running instance of an application a \textit{job}, which consists of a set of \textit{tasks}, each representing a running instance of a microservice. A proper workload placement strategy allows service providers to maximize resource utilization and, more importantly, their revenue. The microservices architecture enables more flexibility in workload placement.

\begin{enumerate}[wide]
    \item \textit{Job Offloading}: IoT edges usually have low computing capacities, and thus a natural solution is to offload workloads to an on-premise edge or an edge cloud. However, doing so naively such as offloading every frame, can incur considerable unnecessary overhead due to network latency. Furthermore, offloading and processing a frame without any object or any target object is a waste of network and computation resources. One promising solution is to leverage a filter to eliminate redundant frames before offloading \cite{FilterForward, BLVA, Clownfish, ESMO}. Moreover, to reduce the communication overhead, frame compression or subsampling can be performed \cite{fastva}. The key is to reduce the amount of transmitted video data during streaming to save bandwidth without compromising accuracy. For instance, CloudSeg \cite{CloudSeg} and Runespoor \cite{Runespoor} stream low-resolution videos, but a super-resolution procedure is applied in the cloud to recover the original high-resolution frames. Similarly, DDS \cite{DDS} continuously sends low-quality videos to an edge server, where an advanced DNN model is employed to determine regions that are likely to be missing. Based on the feedback from the server, the camera re-sends these regions with higher quality for further inference. In VPaaS \cite{VPaaS}, taking advantage of the negligible communication overhead between cameras and an edge server due to their physical proximity, high-quality videos are streamed to the edge server, where they are compressed and sent to the cloud. The cloud runs a heavy model to extract object regions and then sends these region coordinates back to the edge server for classification. The intuition behind this approach is that object localization can be done well on low-quality videos, but accurate classification requires high-quality input. Liu et al. \cite{EAAR} propose a dynamic video encoding technique, which divides each frame into blocks, and applies different encoding schemes to them. In particular, important blocks that are likely to contain objects are encoded with less compression while stronger compression is applied to the other blocks. What is in common among all the aforementioned works is that they treat the pipeline as one non-divisible job. When a frame is offloaded, an edge or cloud server will be responsible for the entire processing. Such approaches are called job offloading (or full offloading).
    
    \item \textit{Task Offloading}: In task offloading (or partial offloading), a task is the smallest unit in offloading. This is beneficial since the components of a VAP can have heterogeneous resource demands. For example, background subtraction is CPU-intensive, while CNN-based object detection is GPU-intensive. Therefore, placing these two components on the same device may result in low execution efficiency since the target executor is not optimized for both. Partitioning the VAP into a collection of tasks allows more flexibility for pipeline execution (e.g., distributing tasks to different executors). Distream \cite{Distream} partitions a VAP among a cluster of multiple cameras and an edge server to maximize the throughput. CEVAS \cite{CEVAS} and LEVEA \cite{LAVEA} investigate partitioning for serverless pipelines, where each component is implemented as a stateless function and executed either at the edge or in the cloud. Lightweight serverless functions are a key enabler for rapid deployment and execution on both the edge and the cloud. More fine-grained partitioning is considered in several works \cite{EdgeML, Neurosurgeon, DADS, EHPDNN, DINA, deepsave,couper, amvp, splitplace,CNNPC,DeepSlicing,OJOS,ASEADS} by splitting neural network layers. Inference terminates at a certain partition point, and the features of the intermediate layer are offloaded to another executor, which will finish the remaining inference. Compression techniques, e.g., lossy and lossless encoding \cite{EHPDNN,Cracking}, DNN-based compression \cite{GRACE,STAC,starfish,dco}, can be used to compress the data and reduce the transmission time in conjunction with layer-wise partitioning. Lossy compression methods usually have a higher compression ratio but may distort important information. AgileNN \cite{AgileNN} finds that only few features contribute the most to the inference and applying heavy compression on them will significantly hurt the performance. Therefore, AgileNN leverages eXplainable AI (XAI) techniques to explicitly learn the importance of features during training. In the inference stage, important features are processed on local devices while less important ones are compressed and offloaded to a server for computation. The final prediction is derived by aggregating the results from both devices. References \cite{ADCNN,EdgeDuet,EagleEye,EdgeFlow} represent another type of task offloading, i.e., spatial offloading, where an input frame or feature map is partitioned into smaller blocks for distributed processing. ADCNN \cite{ADCNN} splits a frame into multiple blocks and offloads them to an edge cluster for parallel processing. Zhou et al. \cite{APE} and EdgeFlow \cite{EdgeFlow} propose a similar approach, but they perform partitioning on the feature maps. In EdgeDuet \cite{EdgeDuet}, blocks containing medium and large objects are processed on IoT devices, while those containing small objects are offloaded to an edge server for accurate detection. Similarly, in EagleEye \cite{EagleEye}, blocks containing large frontal faces are processed by lightweight face recognition models on the mobile device, while the rest of the blocks are sent to the cloud for heavy processing. Therefore, in spatial offloading, the processing of each block can be viewed as a task.
    
    \item \textit{Workload Scheduling}: Apart from offloading, workload scheduling is another important problem that needs to be considered, especially for a system with multiple devices or hosting multiple applications. A scheduler needs to account for a number of constraints, including accuracy requirements, resource availability, budgets, etc. Therefore, a scheduling problem is typically formulated as a constrained optimization problem. Based on where tasks are executed, we divide the scheduling problem into two categories: inter-device scheduling and intra-device scheduling. 
    
    Inter-device scheduling finds the optimal workload placement across different devices, and its objective is to maximize the overall analytics performance. Distream \cite{Distream} and SurveilEdge \cite{SurveilEdge} schedule tasks in an edge cluster by migrating tasks from busy devices to idle ones. Notably, Distream also exploits a long short-term memory network (LSTM) to predict potential incoming tasks in the near future and thus avoid placing tasks on nodes that are going to be busy. When more devices are added to the cluster, Distream is able to increase its throughput nearly linearly. It goes from 489.1 IPS with 6 cameras to 1931.4 IPS with 24 cameras\ \textemdash \ a 3.95$\times$ increase, while maintaining low latency. Other works go beyond simply balancing workloads, and incorporate other criteria, allowing the scheduler to have more choices based on the circumstances. VideoStorm \cite{VideoStorm} places new tasks and migrates existing tasks based on three rules: high utilization, load balancing and lag spreading. The device with the highest average of the three scores will be prioritized for placement. In a cluster of 100 machines, VideoStorm schedules 500 tasks in less than 1 second and 8000 tasks in $\sim$3 seconds. With 1000 machines, VideoStorm is still efficient in decision-making, taking $\sim$1 second for 500 tasks and $\sim$5 seconds for 8000 tasks. Similarly, LAVEA \cite{LAVEA} introduces three task placement strategies, i.e., Shortest Transmission Time First (STTF), Shortest Queue Length First (SQLF) and Shortest Scheduling Latency First (SSLF). Based on the experiments, SSLF has the best overall performance among the three. VideoEdge \cite{videoEdge} applies a greedy heuristic that starts with assigning the configuration and placement with the lowest dominant resource demand (i.e., the maximum ratio of demand to capacity across all resources) to each VAP and greedily considers incremental improvements to the VAPs. When dealing with 40\textendash640 tasks, VideoEdge requires only 0.09\%\textendash3.7\% of the typical time to solve the scheduling problem, which is formulated as a Binary Integer Problem (BIP). The scheduler also considers merging common tasks from different applications on the same stream (e.g., two applications need to detect objects in input frames) to further avoid redundant computation. Similar merging approaches are also explored in \cite{MCDNN,Mainstream}. Nexus \cite{Nexus} and \cite{ASEADS}, on the other hand, focus on shared models that operate on different inputs to increase the resource utilization of the underlying GPU hardware through batching.

    Intra-device scheduling, also known as on-device scheduling, decides the granularity and order of tasks on a single device. It is a well-known problem in real-time systems. Compared to real-time scheduling, on-device scheduling of VA tasks often considers GPU and CPU resources, and takes into account “elasticity” in DL inference by selecting models of different complexity for different parts of the inputs. RT-mDL \cite{RT-mDL} considers a task scheduling problem on a resource-constrained edge device hosting different DL applications, each with a respective deadline. This problem is common in multi-application systems such as autonomous driving \cite{RT-mDL}, where a set of DL tasks with heterogeneous latency requirements (e.g., on-road collision detection, pedestrian tracking, driver speech recognition, etc.) need to be executed concurrently on a resource-limited on-board device. To minimize the overall deadline missing rate, RT-mDL proposes a priority-based task scheduler that divides a DL task into CPU and GPU subtasks and schedules them using separate CPU and GPU task queues, which substantially improves the GPU and CPU temporal utilization. To improve the spatial utilization of the GPU, the scheduler employs a GPU packing strategy to enable parallel execution of DL inferences with priority guarantees. Heimdall \cite{Heimdall} considers a similar problem, i.e., dividing DNN tasks into units and orchestrating them between the GPU and CPU with priorities. Unlike RT-mDL, it allows inference tasks to be scheduled on CPUs. DNN-SAM \cite{DNN-SAM} first splits a DNN task into two sub-tasks: 1) a mandatory sub-task dedicated to a critical portion (e.g., containing target objects) of each image and 2) an optional sub-task for processing a down-scaled image, then executes them independently, and finally merges their results as a single output. To achieve efficient and accurate detection performance, two priority-based scheduling algorithms with different optimization objectives (i.e., minimizing latency or maximizing accuracy) are utilized. DeepQuery \cite{DeepQuery} improves GPU utilization by co-locating delay-critical and delay-tolerant tasks on shared GPUs. The future resource demand for delay-critical tasks is predicted. If more resources are required, the batch-size of delay-tolerant tasks will be reduced by the scheduler to release resources. REMIX \cite{REMIX} presents a different sub-task division scheme, which adaptively partitions an input frame into multiple non-uniform blocks, and assigns each block a proper object detector. Specifically, blocks with dense objects will be processed with an expensive object detector, while the others will be handled with a cheap detector or even ignored. The key idea is similar to spatial offloading, i.e., spatially partitioning the inference task on the input frame into multiple sub-tasks, but they differ in that 1) the blocks in REMIX are non-uniform, and 2) REMIX schedules sub-tasks locally on the same device, rather than offloading them. Slightly different from REMIX, NeuLens \cite{NeuLens} processes the blocks in parallel with different sub-networks of a region-aware convolution (ARAC) supernet, which is an ensemble model. The per-block sub-network selection is controlled by a scheduler, based on application-specific SLOs.

\end{enumerate}

\section{Frameworks and Datasets for Video Analytics}
In this section, we introduce existing open-source frameworks for managing EVA systems, and popular public datasets for VA applications.
\subsection{Container Orchestration Frameworks}
\begin{figure}[t]
\centering
\subfloat[]{
\includegraphics[width=0.12\textwidth]{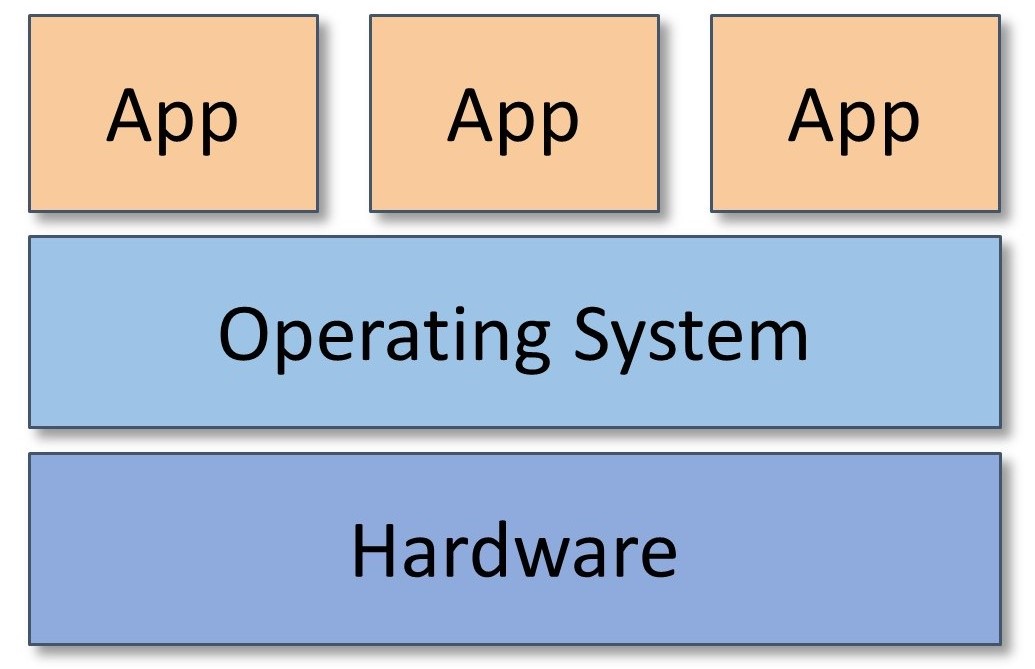}
\label{physical}}
\hfil
\subfloat[]{
\includegraphics[width=0.16\textwidth]{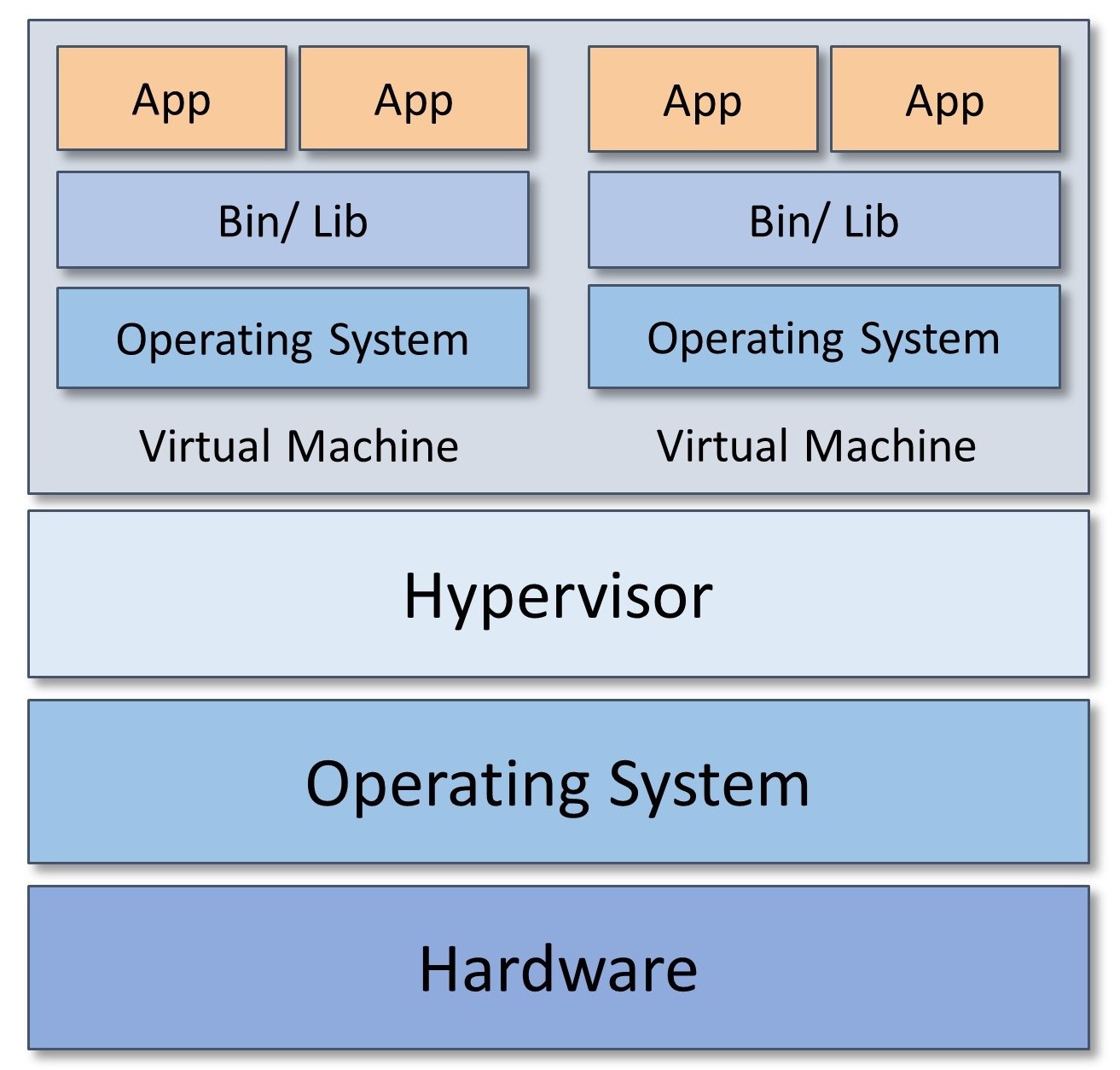}
\label{virtual}}
\hfil
\subfloat[]{
\includegraphics[width=0.16\textwidth]{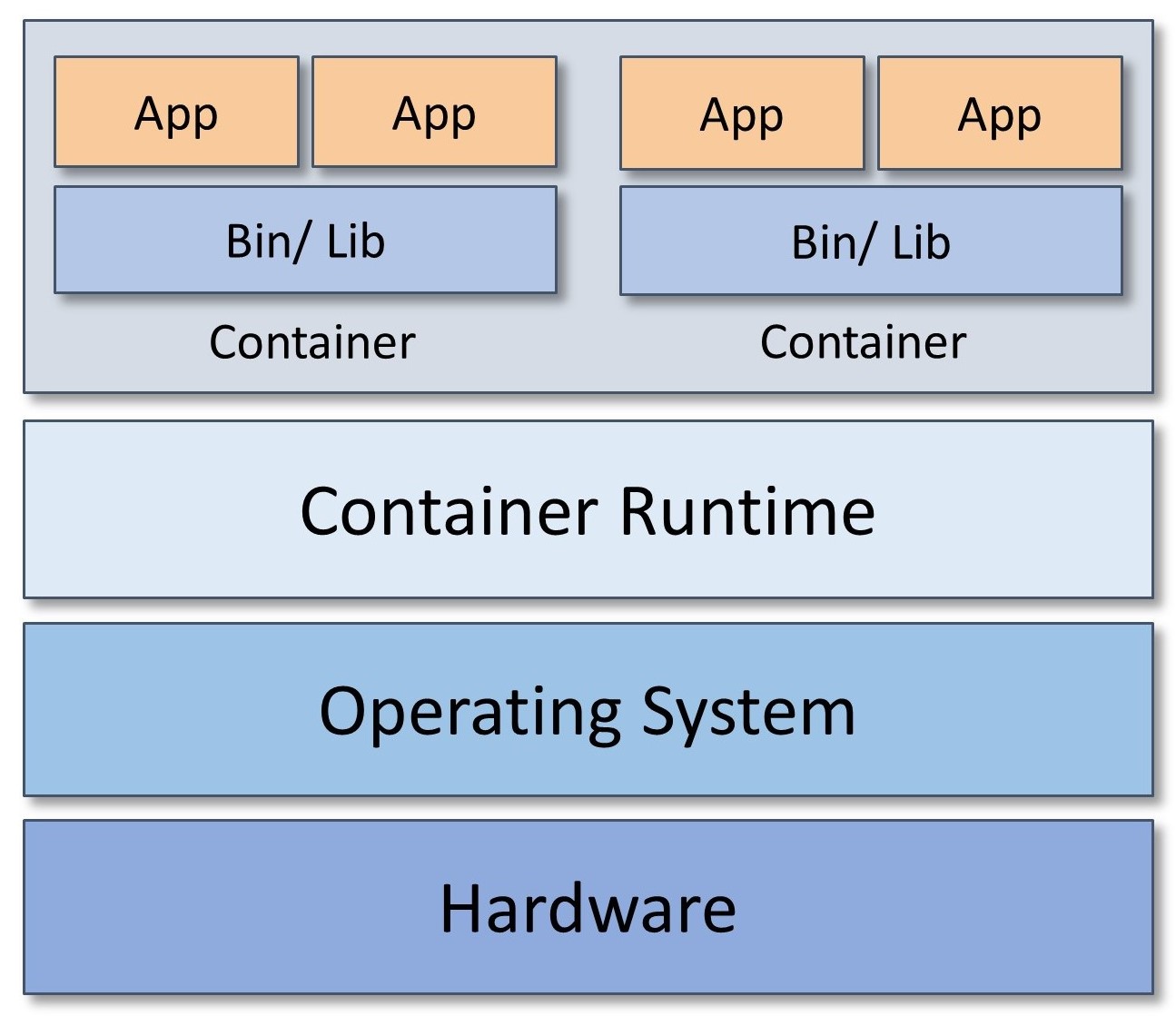}
\label{container}}
\caption{Three ways to deploy an application. (a) Physical machine-based. (b) Virtual machine-based. (c) Container-based.}
\label{virtualization techniques}
\end{figure}

To enable microservices and serverless microservices architecture for applications, one good option is to use containers. Containerization is a new virtualization technique that significantly simplifies and speeds up the creation of isolated containers on VMs or physical machines. Different from VMs, which virtualize hardware to run multiple operating system (OS) instances to host applications, containers encapsulate a lightweight virtualization runtime environment for applications on a single OS, as shown in Fig. \ref{virtualization techniques}. Containers present a consistent software environment, and one can encapsulate all dependencies of a target application as a deployable unit and run it on different devices, e.g., a laptop, a bare metal server, a public cloud, etc.

Many organizations use container orchestration frameworks to manage containers \cite{cncfsurvey}. Orchestration is a way of automating the operational efforts required for managing containerized applications, such as scale-in, scale-out, networking, deployments of containers, etc. All of these operations mentioned above can also be done without an orchestrator if the containerized application to be managed is very small. But when it comes to large-scale applications with hundreds of microservices running thousands of containers, it becomes challenging to manage all these containers, and orchestrators come to the rescue.

\begin{figure}[tb]
\centering
\includegraphics[width=0.46\textwidth]{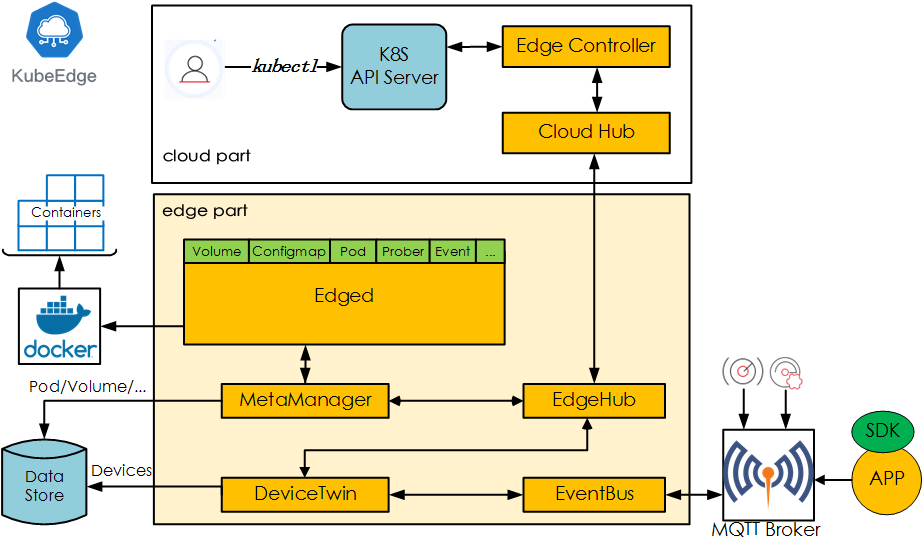}
\caption{Architecture of a KubeEdge cluster \cite{k8skubeedge}.}
\label{kubeedge arch}
\end{figure}

Here, we are going to introduce and compare four widely-adopted container orchestration frameworks:
\begin{enumerate}[wide]
    \item \textit{Kubernetes}: Kubernetes \cite{Kubernetes1,Kubernetes2}, also known as K8s, was initially developed by Google and is currently managed by the Cloud Native Computing Foundation (CNCF). According to the 2021 CNCF Annual Survey \cite{cncfsurvey}, 96\% of organizations are using or evaluating K8s (up significantly from 83\% in 2020 and 78\% in 2019), and more than 5.6 million developers are currently using K8s. By far, K8s has been adopted by major cloud platforms, e.g., Google Kubernetes Engine (GKE), Amazon Elastic Kubernetes Services (EKS) and Microsoft Azure Kubernetes Services (AKS), etc. 

    In general, K8s manages the complete lifecycle of containerized applications in a cluster. It provides high availability, scalability, and predictability to containerized applications, and automates their deployment, management, and scaling. K8s also supports automated rollouts and rollbacks, service discovery, storage orchestration, scaling, batch execution, etc. 

    However, K8s was initially designed to run in cloud environments. Before considering K8s as an orchestrator for EVA systems, there are some major technical challenges to overcome: \begin{itemize}
        \item \textit{Limited Resource at Edge}: Vanilla K8s requires 2 CPUs (cores) and 2 GB memory \cite{k8scpu}. However, edge devices, such as IoT devices, often do not have enough hardware resources to support a complete base K8s deployment. This may limit the applicability in edge computing that contains resource-constrained devices but requires features like high availability, scalability, and fault-tolerance to work in critical areas like surveillance in smart cities \cite{k8scpu}. 
        
        \item \textit{Edge Autonomy}: Vanilla K8s does not have good support for offline independent operations of edge devices (also known as \textit{edge autonomy} \cite{KubeEdge1}). The control plane of a K8s cluster needs to frequently request status information from the nodes to schedule and manage the workloads properly throughout the cluster. Edge computing environments often have restricted connectivity to the Internet in terms of bandwidth and latency, and as a result, the control plane cannot communicate with the edge nodes as much as it needs. Worse still, connectivity can be lost during network outages. The nodes can no longer function without access to the control plane.
        
        \item \textit{Heterogeneous Device Management}: One important feature of edge computing is the distributivity and heterogeneity of edge devices. The devices and their hardware architectures, configurations and communication protocols can significantly vary across application scenarios. Vanilla K8s lacks support for heterogeneous device management and edge-to-edge communication.
       
    \end{itemize} 

    Recently, lightweight K8s distributions have emerged to address the aforementioned challenges and facilitate K8s-based deployments in edge computing settings. Frameworks such as MicroK8s, K3s, KubeEdge provide K8s-compatible distributions by modifying and reorganizing essential components. They aim to simplify configuring, running, and maintaining clusters to enable deployments with low-end edge devices.

    \item \textit{MicroK8s}: MicroK8s \cite{mk8s} is a low-ops, minimal production K8s developed by Canonical. It is an open-source framework for automating the deployment, scaling, and management of containerized applications. MicroK8s aims to solve the challenge of limited edge resources, and provides the functionality of core K8s components, in a small footprint of 564 MB, scalable from a single node to a high-availability production cluster. By reducing the resource commitments required in order to run K8s, MicroK8s makes it possible to run K8s on low-end edge devices, which is beneficial for small-appliance IoT applications.
    \item \textit{K3s}: Rancher offers K3s \cite{k3s} as a lightweight K8s distribution for edge environments, IoT devices, and even ARM devices like the Raspberry Pi. It is fully compliant with K8s, contains all basic components by default, and targets a fast, simple, and efficient way to provide a highly available and fault-tolerant cluster to a set of nodes. The minimum hardware requirements of K3s are 1 CPU and 512 MB of memory, which makes it feasible for edge computing use cases.

    \item \textit{KubeEdge}: KubeEdge \cite{KubeEdge1,KubeEdge2}, was first developed by Huawei, and later accepted as a CNCF sandbox project. Unlike MicroK8s and K3s, which are simply lightweight K8s distributions, KubeEdge is specifically designed to build edge computing solutions by extending the cloud. As shown in Fig. \ref{kubeedge arch}, the KubeEdge architecture consists of cloud, edge, and device layers. In the cloud layer, the K8s API server represents an unchanged native Kubernetes control plane. The CloudCore contains EdgeController and DeviceController, which process data from the control plane, as well as Cloud Hub, which sends the data to EdgeHub at the edge. The edge layer enables application and device management. Specifically, Edged is for application management, whereas DeviceTwin and EventBus are for device management. DataStore facilitates local autonomy. In particular, when the data of an application or a device is distributed from the cloud through EdgeHub, the data is stored in a database before it is sent to Edged or the device. In this way, Edged can retrieve metadata from the database and the service recovers even when the edge is disconnected from the cloud or when the edge node restarts. For device connectivity, KubeEdge supports multiple communication protocols and uses MQTT as a common middleware layer. This helps in scaling the edge clusters with new nodes and devices efficiently. For AI workloads, KubeEdge provides its own toolkit called \textit{Sedna} to make deploying models from popular ML frameworks like Tensorflow and Pytorch easier.
    
    Currently, KubeEdge is gaining popularity due to its lightweight feature (requiring only 66 MB footprint and 30 MB memory) \cite{k8skubeedge} and flexible approach to making edge computing secure, reliable, and autonomous. 
\end{enumerate}

To summarize, although K8s is an industry leader in container management, competitors like K3s, MicroK8s and KubeEdge are viable alternatives in edge computing contexts with their respective strengths and weaknesses. For instance, K3s and MicroK8s are quite mature with extensive documentation and community support, but their functionalities are limited. KubeEdge is an attractive solution due to its features tailored for edge deployments, but it is still in its infancy. Hence, the best framework varies depending on the application requirements and technical competency.

\subsection{Datasets}
Many image or video datasets exist for CV tasks, such as ImageNet \cite{imagenet}, Microsoft Common Objects in Context (COCO) \cite{COCO}, Canadian Institute for Advanced Research (CIFAR) \cite{cifar}, PASCAL Visual Object Classes (VOC) \cite{PascalVOC}. In this section, we limit the discussion to datasets that are widely adopted in 2D MOT tasks:

\begin{enumerate}[wide]
    \item \textit{MOTChallenge}: The goal of MOTChallenge is to provide benchmarks for MOT methods. Several variants were released each year, including MOT15 \cite{MOT15}, MOT16 \cite{MOT16}, MOT17 \cite{MOT16}, MOT20 \cite{MOT20}. Each dataset includes video sequences captured in different places and under different conditions (e.g., camera motion, viewpoint, environment illumination, weather condition, etc.) and the corresponding annotations (e.g., box coordinates, object class, object id, etc.).
    \item \textit{KITTI}: The dataset from Karlsruhe Institute of Technology and Toyota Technological Institute (KITTI) \cite{KITTI} is one of the most popular datasets in mobile robotics and autonomous driving. It consists of hours of traffic scenarios recorded with a variety of sensor modalities, including high-resolution RGB, grayscale stereo cameras, and a 3D laser scanner \cite{KITTI}. KITTI 2D tracking dataset is transformed from 3D data. It consists of 21 training sequences and 29 test sequences. Despite the fact that 8 different classes are labelled, only the classes “Car” and “Pedestrian” are evaluated in the benchmark, because there are not enough labelled instances for other classes.
    
    \item \textit{TAO}: Most MOT benchmarks (e.g., MOT, KITTI) focus on either people or vehicles, motivated by surveillance and self-driving applications. Moreover, they tend to include only a few dozen videos, captured in outdoor or road environments, which may limit the generalizability of models trained using these datasets. To bridge this gap, Tracking Any Object (TAO) \cite{TAO} was proposed. It consists of 2907 high-resolution videos that were captured in diverse environments with an average length of 30 seconds and contain annotations for 833 object categories \cite{TAO}. Different from other datasets which have a limited vocabulary of categories, TAO focuses on diversity both in the category and visual domain distribution, resulting in a realistic benchmark for MOT tasks \cite{TAO}.
    
    \item \textit{BDD100K}: BDD100K \cite{BDD100K} is a large-scale tracking dataset collected from diverse driving scenarios, covering New York, Berkeley, San Francisco Bay Area, and other regions in the US. It contains scenes in a wide variety of locations, weather conditions and day time periods, such as city streets, tunnels, highways, snowy, rainy, cloudy weather, etc. The BDD100K MOT dataset contains 2000 fully annotated 40-second sequences at 5 FPS under different weather conditions, time of the day, and scene types \cite{BDD100K}. The videos contain a total of 130.6K tracking identities and 3.3M objects.
    
    \item \textit{UAVDT}: UAVDT \cite{UAVDT} is a large-scale challenging unmanned aviation vehicle (UAV) Detection and Tracking benchmark for object detection, single object tracking (SOT) and MOT from aerial videos. The objects of interest in this benchmark are vehicles, and it consists of 100 video sequences, which are selected from over 10 hours of videos taken with a UAV platform at a number of locations in urban areas, such as squares, arterial streets, toll stations, highways, crossings and T-junctions \cite{UAVDT}. The videos are recorded at 30 FPS, with a resolution of 1080×540 pixels. The frames are manually annotated with bounding boxes and various attributes, e.g., weather conditions, flying altitude, camera view, vehicle category, etc. 
    
    \item \textit{UA-DETRAC}: UA-DETRAC \cite{UA-DETRAC} is a challenging real-world multi-object detection and multi-object tracking benchmark. It consists of 10 hours of videos captured with various illumination conditions and shooting angles at 24 different locations (e.g., urban highways, traffic crossings and T-junctions) in Beijing and Tianjin, China. The videos are recorded at 25 FPS, with a resolution of 960×540 pixels. More than 140K frames are manually annotated with 8250 vehicles and a total of 1.21M bounding boxes are labelled. UA-DETRAC is now a partner with AI City Challenge \cite{AIC17,AIC18,AIC19,AIC20,AIC21}.
    
    \item \textit{MMPTRACK}: Multi-camera systems have widely been deployed in cluttered and crowded environments, where occlusions of the tracked objects often occur. Datasets for multi-camera multi-object tracking (MCMOT) are quite limited due to data collection and annotation challenges. Multi-camera Multiple People Tracking (MMPTRACK) dataset \cite{MMPTRACK} contains around 9.6 hours of videos, with over half a million frame-wise annotations (e.g., per-frame bounding boxes, person identities, and camera calibration parameters) for each camera view. The annotations are done with the help of an auto-annotation system. The videos are recorded at 15 FPS in five diverse and challenging environment settings., e.g., cafe shop, industry, lobby, office, and retail \cite{MMPTRACK}. This is by far the largest publicly available multi-camera multiple people tracking dataset \cite{MMPTRACK}.
\end{enumerate}

\section{Research Challenges and Future Directions}
Edge video analytics is an active area intersecting with many fields, including VA, CV, DL, and edge computing. Despite significant research efforts, there are still some problems that have not been well addressed or are under-explored yet. In this section, we highlight these problems and outline potential research directions in EVA.

\subsection{Adaptive Configuration Optimization}
As mentioned in Section VI-A, a periodic configuration update is necessary as a configuration can become stale with scene changes \cite{Chameleon,Sight}. An under-investigated aspect of existing configuration optimization methods is how often such configurations change. They all rely on a pre-fixed time interval for updating. For instance, in Chameleon \cite{Chameleon}, fresh online profiling is triggered every $T$ seconds to update the configuration. Similarly, in DeepScale \cite{deepscale}, the optimal resolution is updated every $K$ frames. As reported in DeepScale, the setting of $K$ has a significant impact on the accuracy-latency trade-off. However, an ideal setting requires domain expertise and can vary depending on specific tasks. For parking surveillance, a large time interval is sufficient, while for traffic intersection monitoring, more frequent updates are needed since vehicle movement patterns can change at different times of the day. An improper setting can diminish the benefits brought by periodically updating.

CV techniques can be used to extract features, therefore triggering updates. Take vehicle detection as an example. Low-level features like the number of vehicles, the average size of vehicles, and the average velocity of vehicles generated by an object detector can be used. However, such features may not be sufficient. In continuous videos, consecutive frames have negligible differences in these features, but the detection accuracy of frames can vary significantly due to factors like occlusions and camera parameters \cite{camfluc}. In addition, extracting these features accurately requires a reliable object detector and thus poses a chicken-egg problem\ \textemdash \ a full-fledged object detector with high computation complexity is needed to decide when to trigger profiling or configuration updates, but now the question becomes when to apply such a detector. One possible way out of the dilemma is to utilize cloud resources to decide the best configurations by trading off computation complexity (at the edge) with communication complexity (over networks). 

Prior knowledge is also useful in configuration updates. For example, at a traffic intersection, the prior knowledge could be the peak or off-peak hours on weekdays or weekends, the switching interval of traffic lights, and the vehicle movement patterns (e.g., direction, velocity, etc.) corresponding to the traffic signal. Updates can be simply triggered whenever the traffic signal changes or a rush hour begins. We observe that very few works take prior knowledge into consideration with the exception of Distream\cite{Distream}, where the data captured from a traffic intersection is employed to train an LSTM network to predict vehicle movement patterns in the near future. Note that prior knowledge can be used in conjunction with triggering mechanisms based on image features to improve the efficiency of adaptive configuration optimization. 

\subsection{Multi-Camera Collaborative Video Analytics}
The proliferation of on-camera computing resources has spurred the prospect of massive VA on the camera side. To deliver these promises, however, we must address the fundamental systems challenge of utilizing the on-camera resource in a large camera fleet to run VA applications at scale. Existing multi-camera solutions solely focus on saving computation costs by sharing information across cameras \cite{CrossRoI,ComAI}, or improving throughput by balancing workloads across cameras \cite{Distream}. These solutions often work in isolation. We envision a holistic solution that transforms a group of networked cameras into a compute cluster, called a camera cluster, through which the following benefits can be realized \cite{camcluster}:
\begin{enumerate}[wide]
    \item \textit{Saving Computing Resources}: Different applications sometimes use the same set of vision models. This suggests one can share models, in addition to data, among these applications. For instance, object detection is a common building block in many applications. Instead of frequently loading or unloading DNN models into CPU or GPU memory, it is more efficient to leave the models loaded on specific cameras and route “data” to these devices for several reasons. First, loading a large DNN model into memory from external storage is time-consuming. As an example, loading a ResNet50 model to GPU takes about 10 seconds, which is 100× longer than using it to classify images (50 images per second). Second, many VAPs consist of a cascade of operations, where not all models need to be executed for each frame. Finally, pre-loaded DNNs can batch-process frames from multiple cameras together to further reduce computing time.
    \item \textit{Resource Pooling}: By pooling the resources on cameras, one can process each video stream distributively in the cluster. This is beneficial in two aspects. First, application throughput can be improved if workloads are migrated from overloaded cameras to idle cameras, as cameras in the same cluster often have heterogeneous workloads. In Distream \cite{Distream}, the resources of a camera cluster and an edge server are pooled together, and workloads are balanced across cameras and also between the camera cluster and the edge server. Note that Distream employs a two-stage object detection process, where ROIs are first extracted and then fed into classifiers. The two-stage process makes it possible to improve throughput by load balancing despite increased network transfer time. Second, a camera cluster can run more complex and more accurate models than single-camera solutions. This can be accomplished by splitting a video stream into frame groups and processing them in parallel with multiple cameras. For example, if the current frame has not been finished before the arrival of the next frame (this is common when a heavier model is used), the next frame will be sent to another camera for processing.
    \item \textit{Improving Analytics Quality}: As a camera cluster offers direct access to the video streams of all cameras, one can improve video analytics quality by leveraging information from multiple video streams. Such collaboration has been widely used in MCMOT tasks. As each camera has limited FoVs, objects missed by one camera can potentially be captured by other cameras. By sharing intermediate outputs among the cluster, tracking performance can be improved. Even when cameras do not have overlapping FoVs, collaboration can still be beneficial. Consider a small-scale camera cluster with two cameras A and B, monitoring two different park lots of a shopping mall. They all connect to an edge server located in the mall. Even though the two cameras do not have overlapping FoVs, they share the same type of target objects, i.e., vehicles, and similar video characteristics, i.e., vehicle movement patterns. This means that the optimal configuration for camera A is likely to perform well for camera B \cite{Chameleon}. Moreover, the analytics quality can be further improved if the model is continuously trained with data from both A and B. Very few works consider this setting, and thus collaborative VA for non-overlapping cameras is a promising research direction. 
\end{enumerate}

\subsection{Benchmarks, Datasets and Evaluation Methodology}
Comparing the performance of different VAPs rigorously is non-trivial. First, universally accepted benchmarks are currently lacking. As discussed in Section VII-B, many datasets exist, but they primarily target training and evaluating individual building blocks of a VAP, such as object classification, detection and tracking. High-level application-level metrics such as vehicle counting accuracy and event recognition accuracy are rarely considered. Furthermore, existing datasets are often not comprehensive enough to support a thorough evaluation of a VAP \cite{Clarity}. Yoda \cite{Clarity}, to the best of our knowledge, is the first comprehensive VAP benchmark, covering videos captured in diverse scenarios (stationary or moving cameras; day or night; highway, city or rural streets) with wide-ranging content characteristics and dynamics (object speeds, sizes, and arrival rates). The authors find that a VAP is sensitive to different scenes and its performance can vary significantly based on content features. Second, there is an absence of standardized testbeds. Researchers often report results from their individual setups, which are not easily reproducible by others. As elaborated in Section V-B, an EVA system can be implemented in four distinct architectures. Differences in the chosen architecture, coupled with disparate hardware specifications and network settings can affect the overall accuracy-latency trade-off, making it hard for direct comparisons. In light of these observations, we believe that the establishment of standardized benchmarks and testbeds necessitates a collective and collaborative endeavor from the community.

\subsection{Large-Scale Edge Video Analytics Systems}
Despite the demand for large-scale VA, according to our survey, the majority of reported work considers small-scale deployment (e.g., a few IoT devices and one edge or cloud). To support large-scale VA across a variety of applications, infrastructure must be designed to have the following characteristics: geo-distribution \cite{killerapp,Mutas}, to ensure analytics functionality across cameras, edges, private clusters, and public clouds (not just a central location); multi-tenancy \cite{edgecloudmultitenancy}, to capture and handle many queries per camera as well as queries across multiple cameras; and hardware heterogeneity \cite{Scanner}, to flexibly manage a mix of processing capacities (in CPUs, GPUs, FPGAs, and ASICs) and networks. Organizations managing cameras in a large geographical area can benefit from continuous VA on all of their live feeds, such as counting cars in all intersections for city-wide traffic planning.

As argued by Microsoft \cite{killerapp}, large-scale LVA is the “killer app” of edge computing. A geographically distributed architecture of “public clouds-private edge clusters-cameras” is the only feasible approach to meeting the strict real-time requirements of large-scale LVA. Resource management (e.g., resource provisioning, resource scheduling, and resource monitoring) across the hierarchy of “camera-edge-cloud” will pose significant challenges in such settings. The microservice architecture is an attractive option when implementing large-scale EVA systems due to several advantages. As introduced earlier, microservices provide long-term agility and enable better maintainability in complex, large, and highly-scalable systems by allowing the creation of applications based on many independently deployable services with granular, and autonomous lifecycles. Moreover, microservices can scale out independently, allowing one to scale only the functional area that requires additional processing power or network bandwidth to meet demand.

With serverless architecture, microservices are further decoupled into a series of stateless functions, which can be configured and invoked independently. The same function code can be executed by multiple function instances typically implemented by lightweight containers. Thanks to the lightweight nature of functions, they can scale up or down automatically in milliseconds, leading to rapid and flexible responses. This benefit enables serverless functions to scale and react to fine-grained input workload variations without the need for resource management and monitoring. Moreover, the pay-as-you-go pricing strategy of FaaS (Function-as-a-Service) can ensure no money is wasted on idle resources, thereby reaching high cost-efficiency.

We find the majority of work deploys applications in the monolithic architecture, while only a small portion utilizes microservices and serverless microservices architectures. Further investigation is needed to utilize microservices and serverless microservices to efficiently place and execute VAPs on edge devices, especially in multi-tenant use cases \cite{serverless,serverless2,serverless3,serverless4,serverless5,serverless6,Tetris}. 

\section{Conclusion}
Driving by the flourishing of VA and its stringent latency requirements, there is an imperative need to push the VA frontier from the remote cloud to the proximity of end users. To this end, edge computing has been widely recognized as a promising solution to support computation-intensive DL-driven VA applications in resource-constrained environments. The convergence of VA and edge computing gives birth to the novel paradigm of EVA. 

In this paper, we conduct a comprehensive survey of the recent research efforts on EVA. Specifically, we begin by reviewing the basics of edge computing. We then provide an overview of VA, including its definition, components and application categories. Next, we go through the definition, architectures, components, performance indicators, application architectures and enabling techniques of EVA systems. In addition, to bridge the gap between academia and industry, we introduce a number of frameworks that are widely used in the industry to manage the deployment of VA applications. We also collect various prominent MOT datasets. Finally, we discuss the open challenges and future research directions in EVA. We hope this survey can reflect the recent progress in both academia and industry, evoke growing attention, stimulate wide discussions, and inspire further research ideas in EVA.

\section*{Acknowledgement}
This work is supported by the China Scholarship Council (file No.202108320092), NSERC Discovery Grants, the Canada Research Chair program, and the McMaster Faculty of Engineering's Multidisciplinary PhD Research Support. 

\bibliographystyle{IEEEtran}
\bibliography{ref}

\begin{IEEEbiography}[{\includegraphics[width=1in,height=1.25in,clip,keepaspectratio]{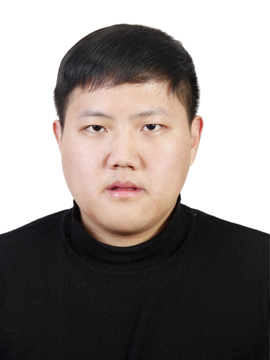}}]{Renjie Xu} (Student Member, IEEE) received his B.E and M.E from Nanjing Forestry University, China, in 2018 and 2021, respectively. He is currently funded by the China Scholarship Council (CSC) and pursuing his Ph.D. degree under the supervision of Dr. Rong Zheng at McMaster University, Canada. His current research interests include video analytics, edge computing, computer vision, and deep learning.
\end{IEEEbiography}

\begin{IEEEbiography}[{\includegraphics[width=1in,height=1.25in,clip,keepaspectratio]{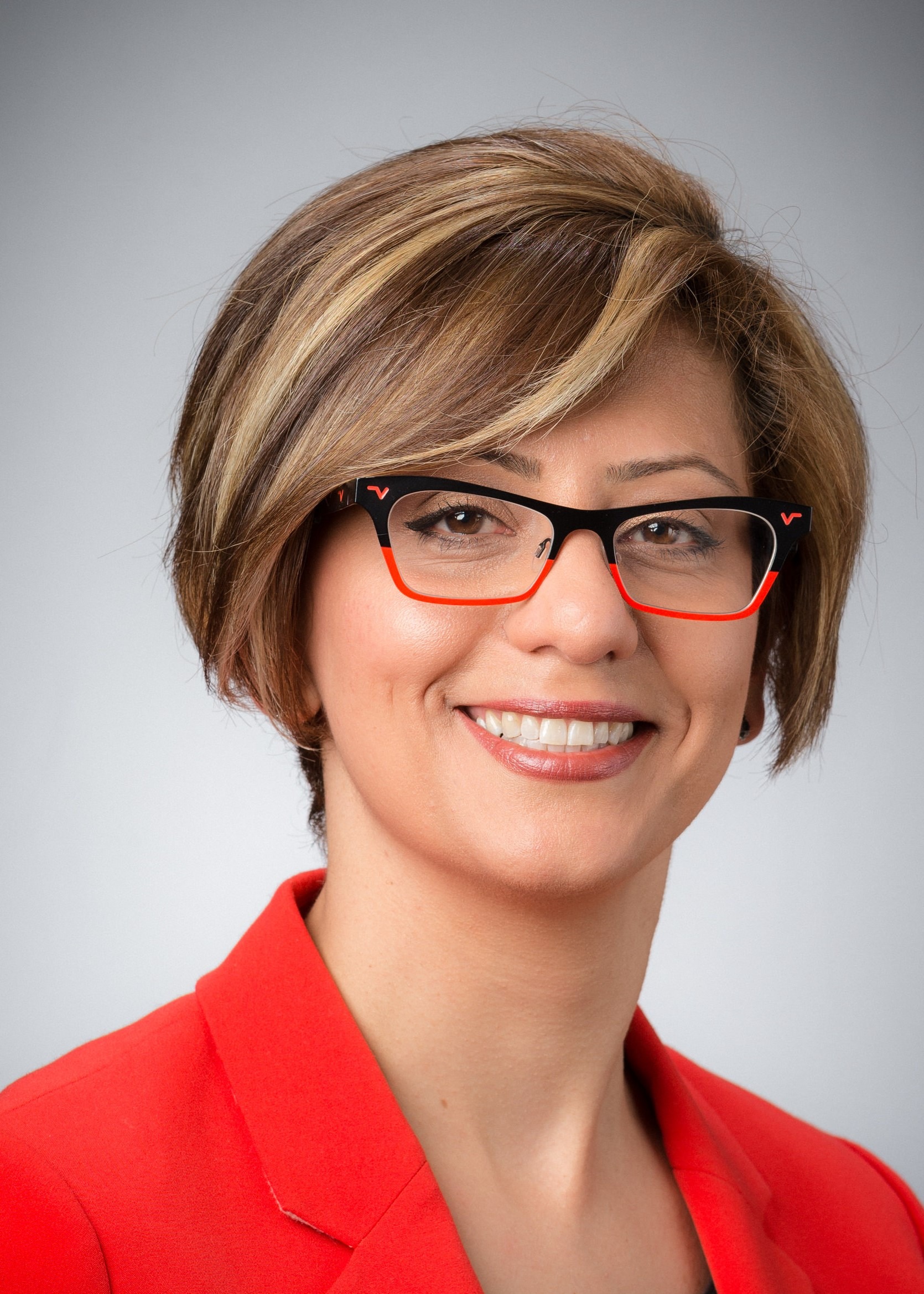}}]{Saiedeh Razavi} (Member, IEEE) is a Professor at the Department of Civil Engineering, the inaugural Chair in Heavy Construction, Co-Director of McMaster’s AI-enhanced Mobility Lab, and Co-Founder of Fluid Intelligence, a multimodal supply chain analytics unit through a strategic alliance with the Hamilton-Oshawa Port Authority. She is also affiliated with the McMaster Institute for Transportation and Logistics (MITL) and the School of Earth, Environment, and Society. Her formal education includes degrees in Computer Engineering (B.Sc), Artificial Intelligence (M.Sc) and Civil Engineering (Ph.D). Her research focuses on AI-enhanced and data-informed knowledge and solutions for smarter construction, intelligent transportation systems, infrastructure management, supply chains and logistics. Dr. Razavi brings together ideas, methods, people, and organizations from different backgrounds and sectors for interdisciplinary and cross-sectoral research to address emerging challenges.
\end{IEEEbiography}

\begin{IEEEbiography}[{\includegraphics[width=1in,height=1.25in,clip,keepaspectratio]{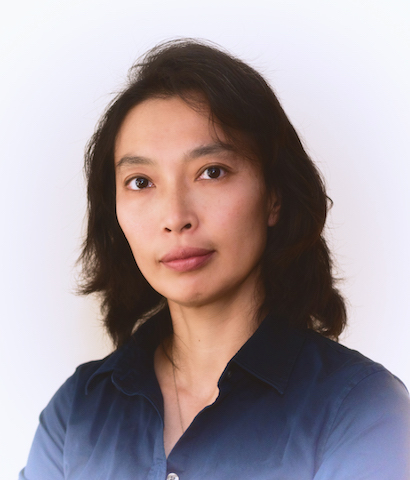}}]{Rong Zheng} (Senior Member, IEEE) received her Ph.D. degree from Dept. of Computer Science, University of Illinois at Urbana-Champaign and earned her M.E. and B.E. in Electrical Engineering from Tsinghua University, P.R. China. She is now a Professor in the Dept. of Computing and Software in McMaster University, Canada. She was on the faculty of the Department of Computer Science, University of Houston from 2004 to 2012. Rong Zheng was a visiting Associate Professor in the Hong Kong Polytechnic University from Aug. 2011 to Jan. 2012; and a visiting Research Scientist in Microsoft Research, Redmond between Feb. 2012 and May 2012.  Rong Zheng’s research interests include mobile computing, data analytics and networked systems. She is currently Tier-1 Canada research chair in Mobile Computing. She was awarded the NSERC Discovery Accelerator Supplement in 2019 and received the National Science Foundation CAREER Award in 2006, and was a Joseph Ip Distinguished Engineering Fellow from 2015 - 2018.
\end{IEEEbiography}

\end{document}